\documentclass[pdflatex,sn-mathphys-num]{sn-jnl}

\usepackage{graphicx}%
\usepackage{multirow}%
\usepackage{amsmath,amssymb,amsfonts}%
\usepackage{amsthm}%
\usepackage{mathrsfs}%
\usepackage[title]{appendix}%
\usepackage{xcolor}%
\usepackage{textcomp}%
\usepackage{manyfoot}%
\usepackage{booktabs}%
\usepackage{algorithm}%
\usepackage{algorithmicx}%
\usepackage{algpseudocode}%
\usepackage{listings}%
\usepackage{siunitx}
\usepackage{upgreek}
\usepackage{makecell}

\theoremstyle{thmstyleone}%
\theoremstyle{thmstyletwo}%

\theoremstyle{thmstylethree}%

\begin{document}

\title[Article Title]{Waveguide-multiplexed photonic matrix-vector multiplication processor using multiport photodetectors}


\author*[1]{\fnm{Rui} \sur{Tang}}\email{ruitang@mosfet.t.u-tokyo.ac.jp}

\author[2]{\fnm{Makoto} \sur{Okano}}

\author[1]{\fnm{Chao} \sur{Zhang}}

\author[1]{\fnm{Kasidit} \sur{Toprasertpong}}

\author[1]{\fnm{Shinichi} \sur{Takagi}}

\author*[1]{\fnm{Mitsuru} \sur{Takenaka}}\email{takenaka@mosfet.t.u-tokyo.ac.jp}

\affil[1]{\orgdiv{Department of Electrical Engineering and Information Systems}, \orgname{The University of Tokyo}, \orgaddress{\postcode{113-8656}, \state{Tokyo}, \country{Japan}}}

\affil[2]{\orgname{National Institute of Advanced Industrial Science and Technology}, \orgaddress{\postcode{305-8568}, \state{Ibaraki}, \country{Japan}}}

\abstract{The slowing down of Moore's law has driven the development of application-specific processors for deep learning. Analog photonic processors offer a promising solution for accelerating matrix-vector multiplications (MVMs) in deep learning by leveraging parallel computations in the optical domain. Intensity-based photonic MVM processors, which do not utilize the phase information of light, are appealing due to their simplified operations. However, existing intensity-based schemes for such processors often employ wavelength multiplexing or mode multiplexing, both of which have limited scalability due to high insertion loss or wavelength crosstalk. In this work, we present a scalable intensity-based photonic MVM processor based on the concept of waveguide multiplexing. This scheme employs multiport photodetectors (PDs) to sum the intensities of multiple optical signals, eliminating the need for multiple wavelengths or modes. A 16-port Ge PD with a 3 dB bandwidth of 11.8 GHz at a bias voltage of -3 V is demonstrated, and it can be further scaled up to handle 250 ports while maintaining a 6.1 GHz operation bandwidth. A 4 $\times$ 4 circuit fabricated on a Si-on-insulator (SOI) platform is used to perform MVMs in a 3-layer neural network designed for classifying Iris flowers, achieving a classification accuracy of 93.3\%. Furthermore, the performance of large-scale circuits in a convolutional neural network (CNN) for Fashion-MNIST is simulated, resulting in a classification accuracy of 90.53\%. This work provides a simplified and scalable approach to photonic MVM, laying a foundation for large-scale and multi-dimensional photonic matrix-matrix multiplication in optical neural networks.}




\maketitle

\section{Introduction}\label{sec1}
The diversity of computational demands in modern applications poses a significant challenge for efficient processing by general-purpose processors such as central processing units (CPUs). This challenge has led to the development and deployment of application-specific processors. For instance, graphics processing units (GPUs) are extensively used in deep learning systems to accelerate matrix operations. However, the slowing down of Moore's law, which traditionally predicted the doubling of transistor density every two years, limits further advancements in energy efficiency for digital electronic processors. Consequently, analog processors based on novel computing mechanisms have garnered significant interest in recent years. Among these, photonic integrated circuits (PICs) emerge as promising platforms for analog processors in deep learning applications. By leveraging the natural advantages of photons in linearity and parallelism, analog photonic processors have the potential to significantly enhance both the computational speed and energy efficiency of deep learning systems, particularly in executing matrix-vector multiplications (MVMs) \cite{shen2017deep, lin2018all, hamerly2019large, wetzstein2020inference, shastri2021photonics, cong2022chip, zhou2022photonic, wang2022optical, ashtiani2022chip, zhu2022space, bernstein2023single, mcmahon2023physics, chen2023deep, chen2023all, xu2024large, xue2024fully, cheng2024multimodal}.

An analog photonic MVM processor performs high-speed and energy-efficient MVMs by propagating light that carries the necessary information. The transfer function of the processor is configured to represent a matrix, and the input light encodes a vector in either its amplitude and phase, or in its amplitude alone. After propagation, the output light automatically carries the information of the matrix-vector product. To date, various circuit architectures have been proposed and demonstrated to realize such photonic MVM processors. Coherent schemes utilize both the amplitude and phase information of light, allowing them to support complex-valued matrices \cite{clements2016optimal, tang2021ten, zhang2021optical, hamerly2022asymptotically, mourgias2022noise, bandyopadhyay2022single, giamougiannis2023coherent, rahimi2024realization, moralis2024perfect}. In contrast, intensity-based schemes support only real-valued matrices since they rely solely on amplitude information \cite{tait2016microring, tait2017neuromorphic, feldmann2021parallel, ohno2022si, zhang2022silicon, dong2023higher, ling2023chip, tang2024symmetric}. One key advantage of intensity-based schemes is their reduced control complexity, as they do not require rigorous phase control or coherent detection. 

MVMs rely on two fundamental operations: element-wise multiplication and addition. In intensity-based schemes, element-wise multiplication is performed by intensity modulations of light, and addition is typically achieved through wavelength multiplexing \cite{feldmann2021parallel, tait2017neuromorphic, ohno2022si, tang2024symmetric}. Optical signals at different wavelengths are multiplexed into a common waveguide, and their total optical power is detected by a single-port photodetector (PD). However, the required number of wavelengths increases with the matrix scale in these schemes. For an $N \times N$ matrix, at least $N$ wavelengths are required. For large values of $N$, the need for efficient multi-wavelength light sources or delicate spectral shaping poses significant challenges for on-chip integration. Moreover, the scalability of wavelength-multiplexed schemes is limited by either high insertion loss or crosstalk from other wavelengths. An intensity-based scheme using mode multiplexing was recently demonstrated \cite{ling2023chip}, in which $N \times 1$ multimode interference (MMI) couplers are employed to combine $N$ optical signals at the same wavelength into a common single-mode waveguide. However, this beam-combining approach introduces a large intrinsic loss that increases with $N$, severely limiting the scalability of this scheme.

In this work, we present a scalable intensity-based photonic MVM processor based on the concept of waveguide multiplexing \cite{tang2022two, tang2024single}. In this approach, the multiplication results encoded in optical intensities within each waveguide are summed using multiport PDs, eliminating the need for wavelength multiplexing or mode multiplexing. A 16-port Ge PD with a 3 dB bandwidth of 11.8 GHz at a bias voltage of -3 V is demonstrated, and it can be further scaled up to handle 250 ports while maintaining a 6.1 GHz operation bandwidth. A 4 $\times$ 4 circuit fabricated on a Si-on-insulator (SOI) platform is used to perform MVMs in a 3-layer neural network designed for classifying Iris flowers, achieving a classification accuracy of 93.3\%. Furthermore, the performance of large-scale circuits in a convolutional neural network (CNN) for Fashion-MNIST is simulated, resulting in a classification accuracy of 90.53\%.

\section{Results}\label{sec2}

\subsection{Principle}\label{subsec1}
Figure 1 shows the schematic structure of a $4 \times 4$ waveguide-multiplexed circuit for photonic MVM. A single-wavelength input optical signal is equally split and then modulated by an intensity modulator array to generate a vector $\boldsymbol x$. The vector is further split into multiple copies, with each copy routed to a modulator array for one row of a matrix $\mathrm{\bf W}$. The intensity modulators can be Mach-Zehnder interferometers (MZIs), microring resonators (MRRs), or phase change material (PCM) absorbers \cite{wuttig2017phase, zhou2023memory, miyatake2024photonic}. The original structure, as proposed in our previous work \cite{tang2022two}, assumes the use of multiple waveguide layers to avoid waveguide crossings. However, small-scale circuits, such as the one shown here, can also be easily fabricated using a single waveguide layer if low-loss and low-crosstalk waveguide crossings are used. The double intensity modulations perform the multiplications between each vector and matrix element. These twice-modulated optical signals are then summed by the multiport PD, generating a photocurrent proportional to the total power of the incident light. The multiport PD can be implemented using Ge-on-Si with a vertical p-i-n structure \cite{siew2021review}, which is compatible with standard complementary metal-oxide-semiconductor (CMOS) process in photonics foundries. While it is also possible to first detect each optical signal with a single-port PD and then combine the photocurrents of these PDs \cite{ashtiani2022chip}, the use of multiport PDs reduces the dark current and capacitance, as the total area of Ge regions is reduced. Therefore, the multiport PD brings significant advantages in signal-to-noise ratio (SNR) and operation bandwidth, compared to a single-port PD array. A comparison between the multiport PD and the single-port PD array is provided in Appendix B.
\begin{figure}[t]
\centering
\includegraphics[width=1.0\textwidth]{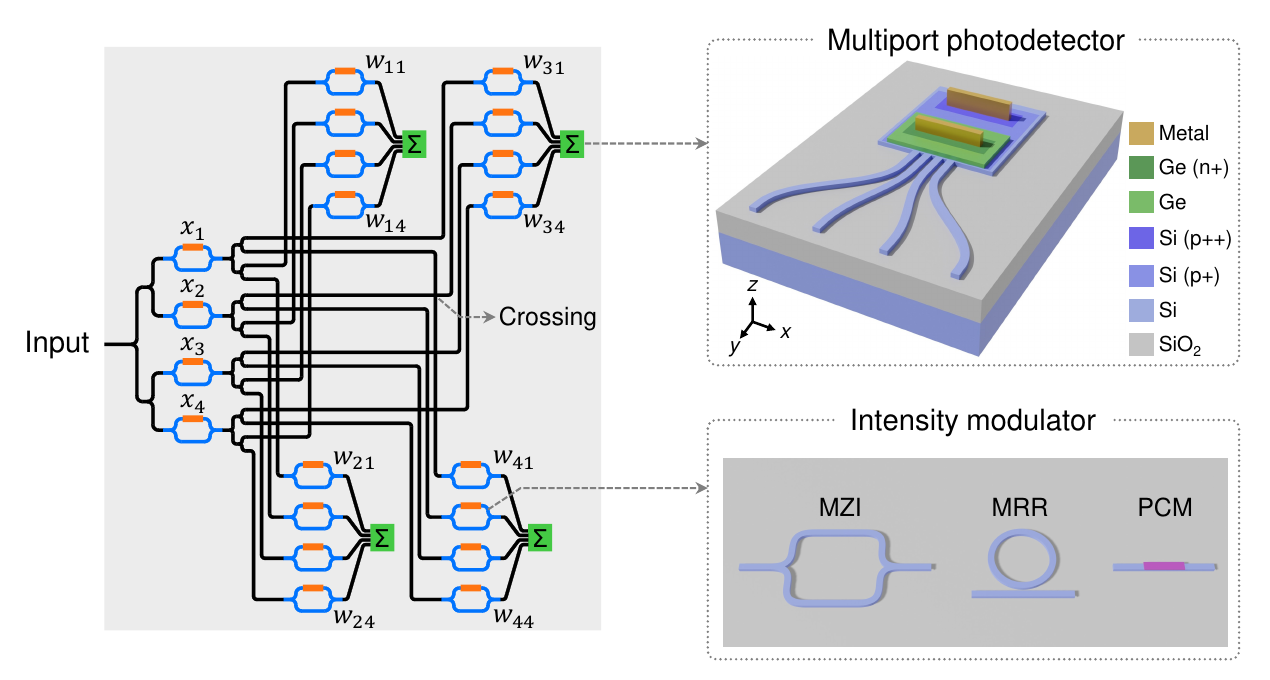}
\caption{Schematic structure of a $4 \times 4$ waveguide-multiplexed circuit using a single-waveguide-layer platform. Intensity modulators, such as MZIs, MRRs, or PCM absorbers, are used to generate the vector and matrix elements. The multiport PD sums multiple optical signals at the same wavelength and can be implemented using Ge-on-Si with a vertical p-i-n structure. Redundant waveguide crossings (not shown in the figure) are inserted into paths with fewer crossings to equalize insertion loss. MZI: Mach-Zehnder interferometer. MRR: microring resonator. PCM: phase change material.}\label{fig1}
\end{figure}

\subsection{Multiport photodetector}\label{subsec2}
\begin{figure}[t]
\centering
\includegraphics[width=1.0\textwidth]{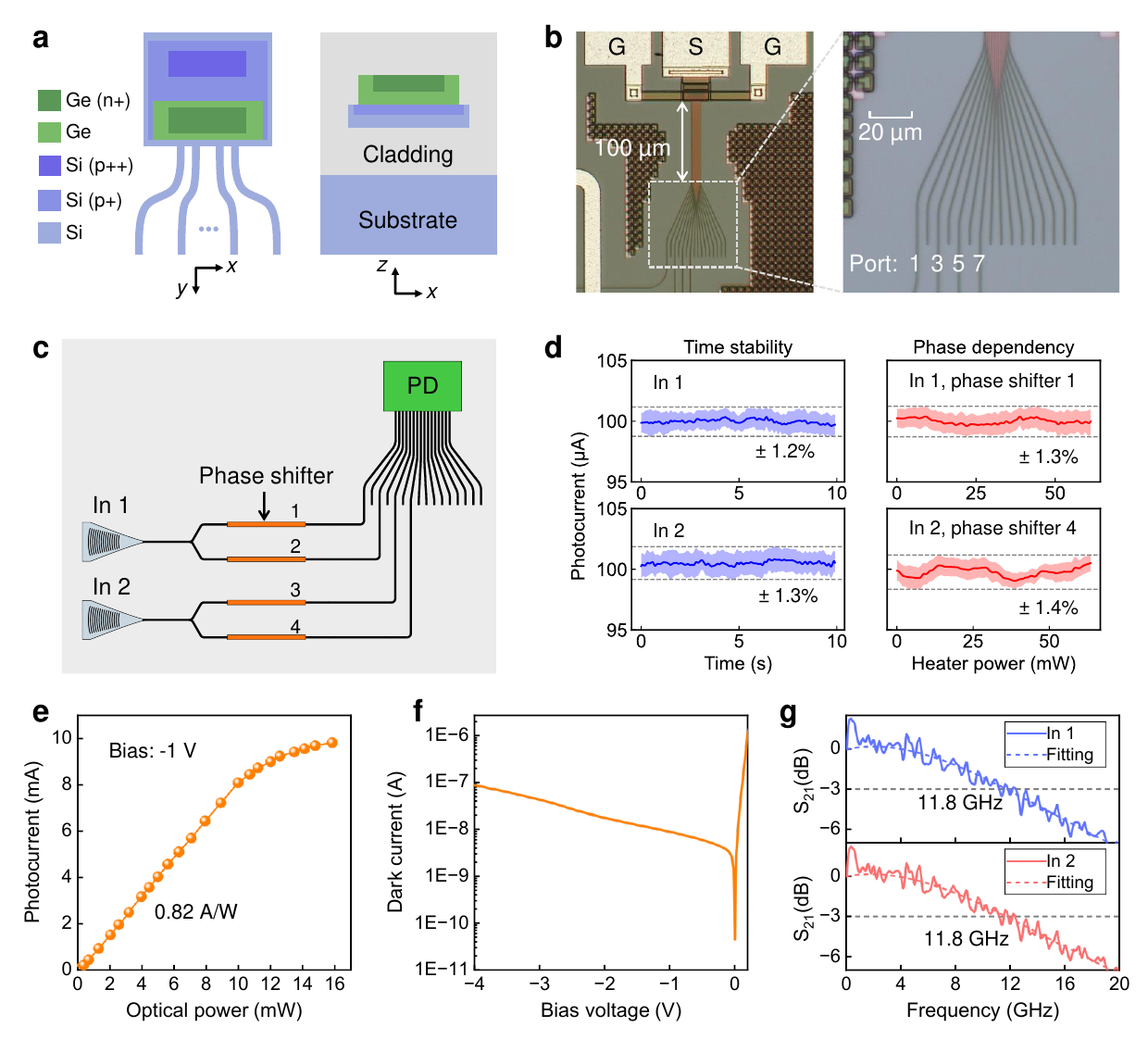}
\caption{\textbf{a} Schematic diagram of the waveguide-coupled Ge-on-Si multiport PD using a vertical p-i-n structure. The drawing is not to scale, and metal layers are not shown. \textbf{b} A fabricated 16-port PD consisting of a 100-\unit{\micro\meter}-long multiport waveguide coupler region. The 300-nm waveguide spacings are nearly invisible. \textbf{c} Four PD ports are connected to on-chip components for further characterization. \textbf{d} Measured time stability of the photocurrent (blue curves) and the response of the photocurrent when the light phase in one path is tuned from 0 to more than $2\uppi$ (red curves). \textbf{e} Photocurrent as a function of the incident optical power. \textbf{f} Dark current as a function of the bias voltage. \textbf{g} Measured electro-optic frequency response.}\label{fig2}
\end{figure}

A schematic diagram of the waveguide-coupled Ge-on-Si multiport PD using a vertical p-i-n structure is shown in Fig. 2(a). Multiple optical signals injected into the PD are simultaneously absorbed in the Ge region. Within its linear operation range, the multiport PD generates a photocurrent proportional to the total power of the incident light, thereby performing a summation via the photoelectric effect. Since the device width is proportional to the number of PD ports, the input waveguides are brought close to each other near the incident interface to achieve a compact device size. Although light coupling and interference can occur among these waveguides, the generated photocurrent should in principle remain unaffected, as long as the light from each port is equally absorbed. To verify this, we designed a 16-port PD consisting of a 100-\unit{\micro\meter}-long multiport waveguide coupler region, where the 16 waveguides have widths of 440 nm and spacings of 300 nm. Light injected into any of the input ports experiences strong coupling in the multiport waveguide coupler \cite{tang2021ten}. Figure 2(b) shows microscope images of a 16-port PD fabricated by Advanced Micro Foundry (AMF), where the 300-nm waveguide spacings are nearly invisible. Among the 16 input ports, 4 ports are connected to other on-chip components for further characterization, as illustrated in Fig. 2(c). Light is coupled into the chip via one of the two grating couplers and further split equally into two paths. Each path has a thermo-optic phase shifter made of a thin-film metal heater, with a power consumption of less than 25 mW/$\uppi$. 

The time stability of the photocurrent is first measured when light at a wavelength of 1550 nm is injected into the two grating couplers separately. Then, the response of the photocurrent is characterized when the phase of light in one path is tuned. The results are shown in Fig. 2(d). Each curve represents the mean value of 30 repeated measurements, with the shaded color band indicating the standard deviation. The measured photocurrent exhibits temporal fluctuations of up to $\pm1.3\%$ over 10 seconds at 100 \unit{\micro\ampere}. This fluctuation is primarily due to laser power instability and the thermal noise of the PD. When the electric power applied to one phase shifter is increased from 0 to more than 60 mW, corresponding to a phase change from 0 to more than $2\uppi$, the fluctuation slightly increases to up to $\pm1.4\%$. Since the light from one grating coupler is injected into two PD ports simultaneously, shifting the light phase in one path significantly alters the light field distribution in the multiport waveguide coupler region. The response of the photocurrent confirms that light coupling and interference among the input ports only slightly affect the operation of the multiport PD. Moreover, in the simulation of large-scale circuits presented in a later section, it is numerically confirmed that adding Gaussian noise with a standard deviation of 1.5\% to the photocurrents does not significantly degrade the circuit performance. On the other hand, the interference-induced fluctuation can be reduced by improving the spatial uniformity of Ge and the uniformity of all PD ports.

Figure 2(e) shows the photocurrent as a function of the incident optical power when light is injected into the grating coupler `In 1'. The photocurrent demonstrates high linearity up to approximately 8 mA, with a responsivity of 0.82 A/W within the linear range. No significant difference is observed when switching the input to `In 2'. Figure 2(f) shows the measured dark current as a function of the bias voltage. The dark current is 43 nA at a bias voltage of -3 V. Figure 2(g) shows the measured bandwidths of the PD when modulated light is injected into the coupler `In 1' and `In 2', respectively. A 3 dB bandwidth of 11.8 GHz at a -3 V bias voltage is observed. The bandwidth is primarily limited by the RC constant and can be improved by optimizing the design of the PD. In Appendix C, we present the results of a large Ge PD, which shows a 3 dB bandwidth of 6.1 GHz at a -3 V bias voltage and can support approximately 250 input ports. These results indicate that the multiport PD is sufficiently scalable.

\subsection{Circuit operation}\label{subsec3}
\begin{figure}[b]
\centering
\includegraphics[width=1.0\textwidth]{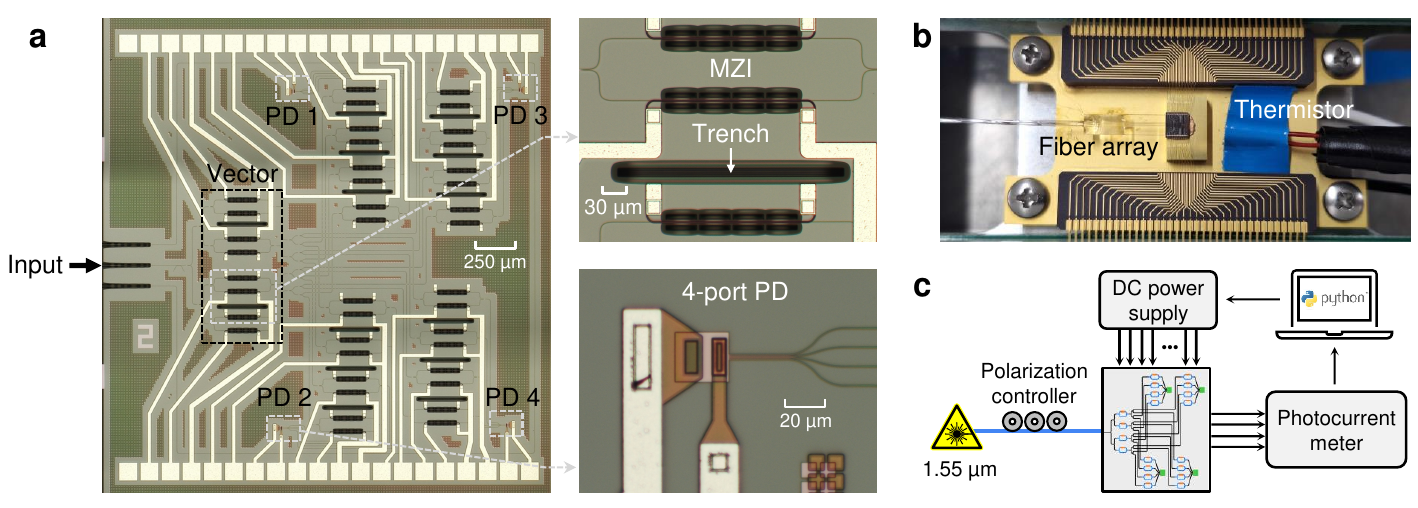}
\caption{\textbf{a} A $4 \times 4$ circuit fabricated on an SOI platform, with enlarged images of an MZI and a 4-port PD. \textbf{b} The circuit is wire-bonded and packaged with a fiber array. \textbf{c} Experimental setup.}\label{fig3}
\end{figure}

Figure 3(a) shows a $4 \times 4$ circuit fabricated on an SOI platform by AMF, which includes 20 tunable MZIs as intensity modulators and four 4-port PDs. The circuit occupies an area of $2.8 \times 3.1$ \unit{mm^2}. Each MZI uses a thermo-optic phase shifter to modulate the light intensity, with a low power consumption of approximately 1.3 mW/$\uppi$. Deep trenches are inserted between MZIs to provide thermal isolation. The 4-port PD has a 30-\unit{\micro\meter}-long multiport waveguide coupler region, with waveguide widths and spacings identical to those of the 16-port PD. Waveguide crossings with low insertion loss and low crosstalks are employed \cite{ma2013ultralow}, and redundant crossings are inserted into paths with fewer crossings. The circuit is then wire-bonded and packaged with a fiber array, as shown in Fig. 3(b). Figure 3(c) illustrates the experimental setup. Light at a wavelength of 1550 nm is injected into the chip with the polarization adjusted to excite the transverse-electric (TE) mode of the Si waveguide. A computer controls a multi-channel direct-current (DC) power supply (Nicslab, XPOW), which drives all the phase shifters in current source mode. The photocurrents of the 4-port PDs are measured using photocurrent meters (Thorlabs, PDA8000).

\begin{figure}[b]
\centering
\includegraphics[width=1.0\textwidth]{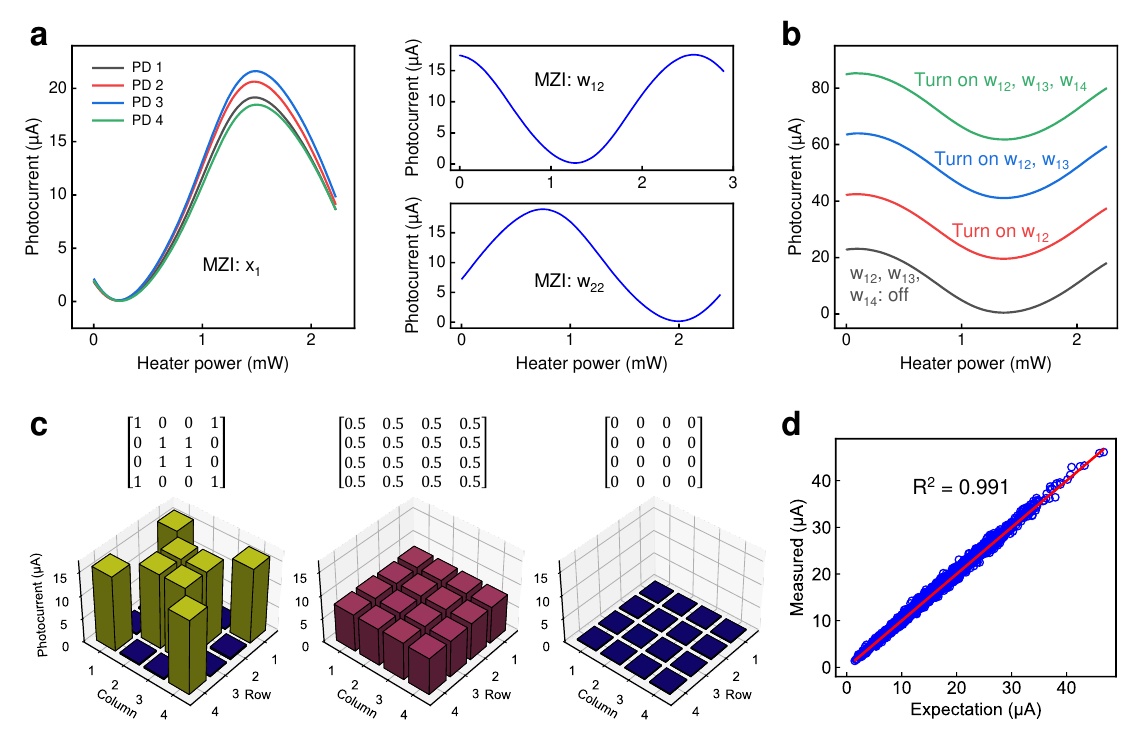}
\caption{\textbf{a} Characterizations of an MZI for a vector element ($x_{1}$) and two MZIs for two matrix elements ($w_{12}$ and $w_{22}$) \textbf{b} Photocurrent response of one 4-port PD (PD-1) when tuning the MZI for $w_{11}$ under various conditions. \textbf{c} Implemented matrices. \textbf{d} Expected and measured photocurrents of the 4-port PDs when 500 randomly generated matrices and vectors are implemented.}\label{fig4}
\end{figure}

The MZIs are characterized by tuning the phase shifters and observing the changes in the photocurrents of the PDs. Figure 4(a) shows the results of characterizing an MZI for a vector element ($x_{1}$) and two MZIs for two matrix elements ($w_{12}$ and $w_{22}$). Irrelevant MZIs are set to the minimum-transmittance state during the characterization. When tuning the MZI for $x_{1}$, the maximum photocurrents of the 4 PDs are slightly different due to non-identical insertion losses among all paths. This difference is then compensated by calibrating the MZIs for the matrix to obtain identical maximum photocurrents. The MZIs for the matrix exhibit different initial phases as a result of fabrication inaccuracies. The measured extinction ratios of all MZIs are greater than 19.5 dB and can be further improved by inserting polarization filters or adding extra stages \cite{tang2024symmetric, wilkes201660}. Figure 4(b) shows the photocurrent response of PD-1 when tuning the MZI for $w_{11}$ under various conditions. Initially, the four vector-MZIs ($x_1 \sim x_4$) are set to the maximum-transmittance state, and the phase shifter in the MZI for $w_{11}$ is tuned while the MZIs for $w_{12}$, $w_{13}$, and $w_{14}$ are set to the minimum-transmittance state. Next, the MZIs for $w_{12}$, $w_{13}$, and $w_{14}$ are sequentially set to the maximum-transmittance state, and the MZI for $w_{11}$ is tuned again. We can observe that the 4-port PD successfully generates photocurrents proportional to the total incident optical power, even though the 4 incident optical signals have different amplitudes and phases.

After characterizing all the MZIs, the matrix-MZIs are calibrated to ensure that the maximum output power from each MZI is nearly the same. These MZIs are then programmed to implement various matrices, as shown in Fig. 4(c). Here, an ideal `0' in the matrix cannot be physically implemented because the extinction ratios of MZIs are not infinitely large. Error signals in these matrices are suppressed to approximately -16 dB. Although this circuit only supports positive matrix elements, negative values can also be implemented if $1 \times 2$ MZIs are used, as discussed in our previous work \cite{tang2022two}. We further implement 500 randomly generated matrices and vectors, and compare the expected and measured photocurrents of the 4-port PDs, as shown in Fig. 4(d). The results show a high determination coefficient ($R^2$) of 0.991, indicating a high operational fidelity of the circuit.

\subsection{Classification of Iris flowers}\label{subsec4}
The classification of Iris flowers is a common task performed by small-scale artificial neural networks. Here, we construct a 3-layer neural network to classify Iris flowers, as illustrated in Fig. 5(a). The open-source dataset contains 150 samples, each characterized by 4 features (sepal length/width and petal length/width) and labeled with the corresponding flower type \cite{iris}. We randomly divide the dataset into two parts: 90 samples for training and 60 samples for testing. The input vector is normalized between 0 and 1 for physical implementation. The sigmoid function, which outputs values between 0 and 1, is used as the nonlinear activation function in the hidden layer. In the output layer, the item with the highest value is selected as the predicted result. This neural network is trained on a computer using a stochastic gradient descent algorithm, with the weight matrices constrained to be non-negative. After training, a classification accuracy of 95\% is achieved on the test dataset using the computer alone, as shown in Fig. 5(b). We then use the $4 \times 4$ circuit to experimentally perform the two MVMs in this neural network, with the trained weight matrices normalized between 0 and 1. Other operations are still performed by the computer. For the same test data, a classification accuracy of 93.3\% is achieved. The slight decrease in classification accuracy is attributed to hardware errors in the circuit, including the finite extinction ratios of MZIs, temporal fluctuations of photocurrents, and crosstalks caused by the waveguide crossings. 

For an $N \times N$ circuit, the compute speed in operations per second (OPS) is given by $2fN^2$, where $f$ is the clock frequency of the circuit. For proof-of-concept demonstrations, thermo-optic phase shifters are used in the $4 \times 4$ circuit. Since these phase shifters use deep trenches for thermal isolation, the modulation bandwidth is limited to a few kHz \cite{gao2022thermo}. Assuming a clock frequency of 3 kHz, the compute speed of this circuit is estimated to be $9.6 \times 10^4$ OPS. The compute speed can be significantly increased by using a larger-scale circuit and intensity modulators with GHz-level bandwidths \cite{moralis2024perfect}. For instance, with a matrix size of $64 \times 64$ and a clock frequency of 1 GHz, the estimated compute speed can reach up to 8.2 tera-OPS.
\begin{figure}[t]
\centering
\includegraphics[width=0.75\textwidth]{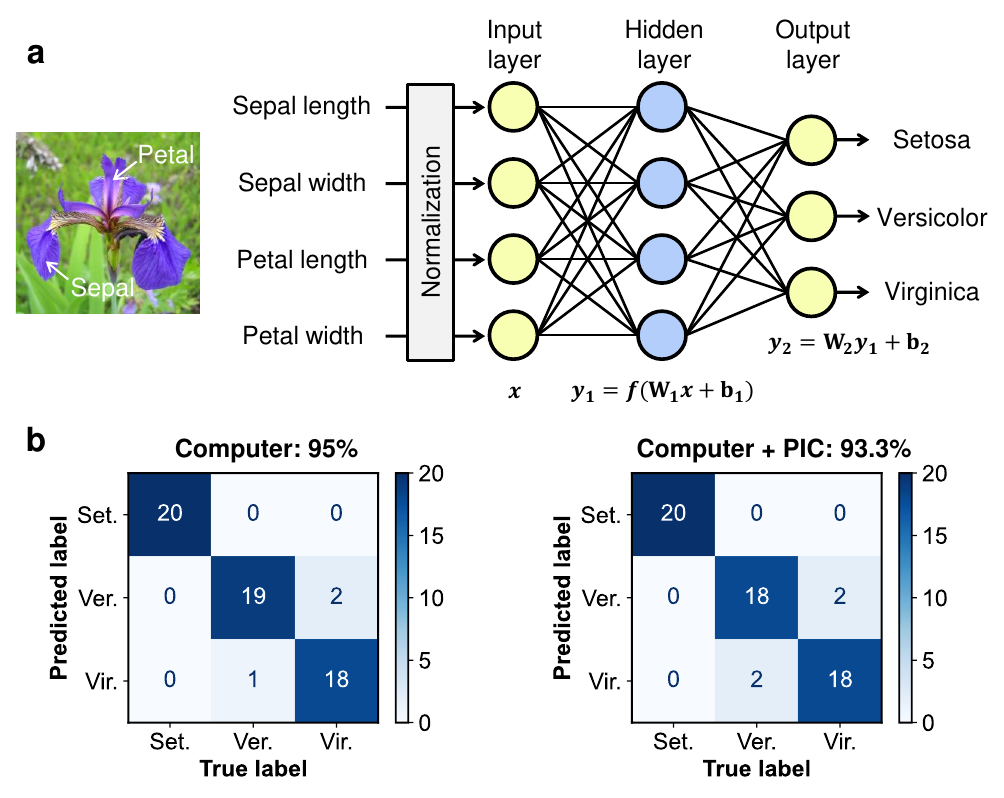}
\caption{\textbf{a} A 3-layer neural network for classifying Iris flowers. \textbf{b} Classification results when a computer alone is used (left) and when the MVMs are performed by the $4 \times 4$ circuit (right).}\label{fig5}
\end{figure}

\subsection{Simulation of large-scale circuits}\label{subsec5}
\begin{figure}[t]
\centering
\includegraphics[width=0.9\textwidth]{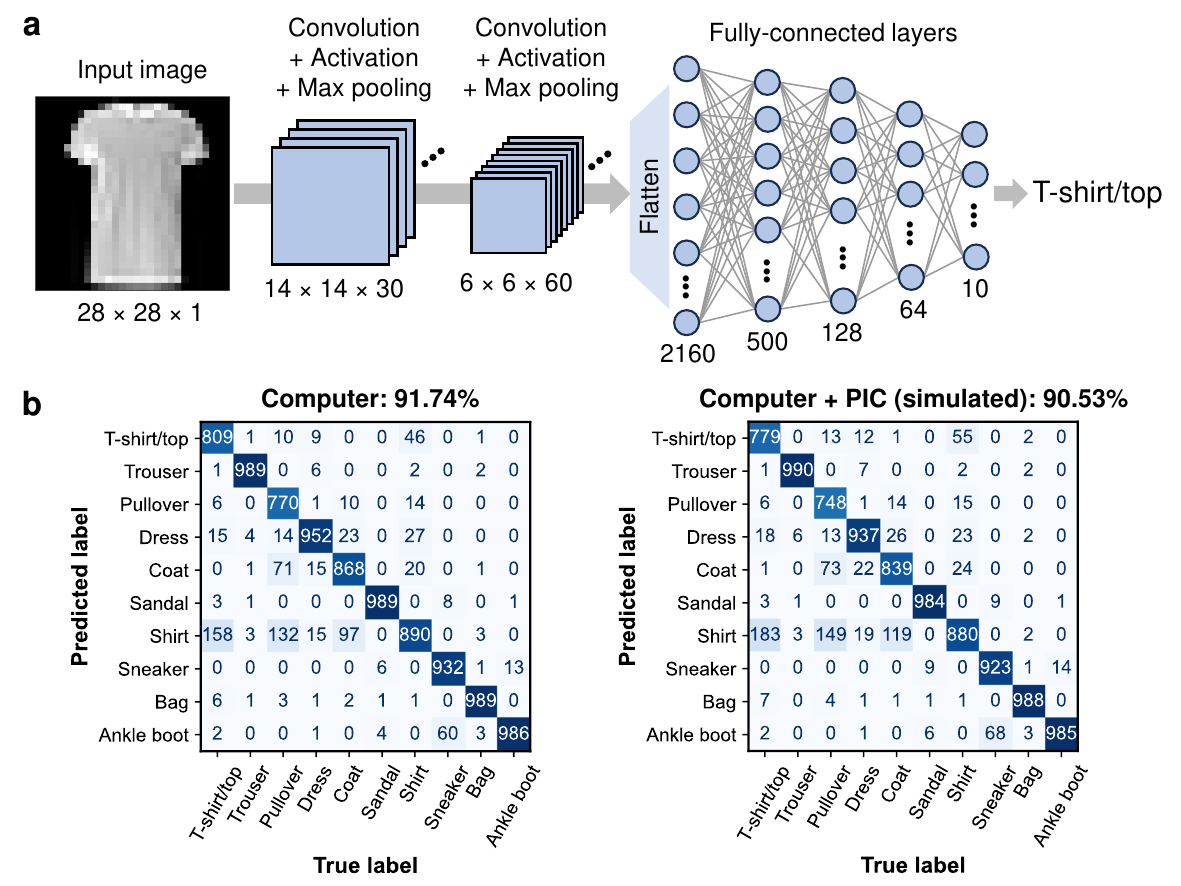}
\caption{\textbf{a} A CNN for Fashion-MNIST consisting of two convolutional layers and several fully-connected layers. \textbf{b} Classification results when a computer alone is used (left) and when the MVMs in the last two layers are calculated using a developed model (right).}\label{fig6}
\end{figure}

We simulate the performance of large-scale circuits when used in a CNN for Fashion-MNIST, which is a dataset of Zalando's article images consisting of a training set of 60000 examples and a test set of 10000 examples \cite{fashionMNIST}. The network architecture is shown in Fig. 6(a). An input image is processed through two convolutional layers followed by several fully connected layers. Large-scale circuits are assumed to perform the MVMs in the last two layers, where the matrix sizes are $64 \times 128$ and $10 \times 64$, respectively. Note that non-square matrices, as used here, are also supported in this scheme either directly or with slight adaptations: $64 \times 128$ matrices are directly supported in a $64 \times 128$ circuit, while $10 \times 64$ matrices are supported by replacing the multi-stage splitters with directional couplers, as introduced in the original proposal \cite{tang2022two}, or by partially using a $16 \times 64$ circuit.

A model is developed to simulate the circuit output, taking into account the finite extinction ratio of MZIs, fluctuations in photocurrents, and finite control/readout resolutions. We assume that the extinction ratio of all MZIs is 30 dB, the fluctuations in photocurrents follow normal distributions with a standard deviation of 1.5\%, the transmittances of MZIs are controlled with an 8-bit resolution, and the photocurrents of multiport PDs are sampled with an 8-bit resolution. Since the use of waveguide crossings is not suitable for large-scale circuits, we assume these circuits use multiple waveguide layers and are therefore crossing-free. This neural network is again trained on a computer, and a classification accuracy of 91.74\% is achieved on the test dataset using the computer alone, as shown in Fig. 6(b). Next, the MVMs in the last two layers are calculated using the developed model, with other operations remaining unchanged. In this case, the classification accuracy remains high at 90.53\%, suggesting that high-fidelity MVMs are achievable with large-scale circuits.

\section{Discussion}\label{sec3}
\subsection{Comparison with previous works}\label{subsec3_1}
In Table 1, we compare this scheme with previous works. Here, we focus on on-chip schemes that can perform MVMs within one modulation cycle; thus, schemes based on time-domain multiplexing or continuous-time data are not included in this comparison \cite{hamerly2019large, xu202111, sludds2022delocalized, dong2023higher, rahimi2024realization, chen2024hypermultiplexed}. This work stands out as an intensity-based scheme that uses a single wavelength and a single waveguide mode for MVMs, enabled by the concept of waveguide multiplexing. Furthermore, while the wavelength dimension or mode dimension is not necessary for MVM in this scheme \cite{mojaver2024recent}, they can be introduced to enable matrix-matrix multiplications, further improving computation speed and energy efficiency. The compatibility with these two dimensions is discussed in Appendix D.

\begin{table}[h]
\caption{Comparison with previous works}\label{tab1}%
\begin{tabular}{@{}llllll@{}}
\toprule
Ref. & Year & Circuit scale & Coherence & Wavelength & Waveguide mode\\
\midrule
\cite{shen2017deep} & 2017 & $26 \times 6$ & Coherent & Single & Single \\
\cite{tait2017neuromorphic} & 2017 &$4 \times 4$ & Intensity-based & 4 & Single \\
\cite{zhang2021optical} & 2021 &$6 \times 6$ & Coherent & Single & Single \\
\cite{feldmann2021parallel} & 2021 &$16 \times 16$ & Intensity-based & 16 & Single \\
\cite{ling2023chip} & 2023 &$4 \times 4$ & Intensity-based & Single & 4 \\
\cite{ohno2022si, tang2024symmetric} & 2022, 2024 & $4 \times 4$ & Intensity-based & 4 & Single \\
\cite{moralis2024perfect} & 2024 & $4 \times 4$ & Coherent & Single & Single \\
This work & 2024 & $4 \times 4$ & Intensity-based & Single & Single \\
\botrule
\end{tabular}
\end{table}

\subsection{Scalability}\label{subsec3_2}
The scalability of this scheme is primarily determined by two factors: the multiport PD and the on-chip insertion loss. Increasing the number of ports in the multiport PD results in higher dark current and slower response speed. In Appendix C, we present the measured frequency response of a large Ge PD (Ge area: $200 \times 8$ \unit{\micro\meter^2}), which shows a 3-dB bandwidth of approximately 6.1 GHz at a bias voltage of -3 V. With a port pitch as small as 0.74 \unit{\micro\meter} at the incident interface, this PD can support approximately 250 input ports. The measured dark current at -3 V is 127.7 nA. Since the operation bandwidth and dark current depend solely on the Ge area and p-i-n structure, these results indicate that the multiport PD can scale up to handle hundreds of ports while maintaining GHz-level operation speed.

The on-chip loss in this scheme is primarily attributed to waveguide propagation loss and the extra loss of optical components. In Appendix E, we show the estimated insertion loss and chip area for this scheme, assuming that MZIs and PCM absorbers are used as intensity modulators separately. Another source of loss, the matrix loss, results from intensity modulations. Since it is universal across all intensity-based schemes, matrix loss is not considered in this estimation. To reduce chip area and propagation loss, we have assumed a slightly modified structure, in which light is split using directional couplers instead of multi-stage optical splitters. Since a PCM modulator is significantly more compact than an MZI modulator, implementing this scheme with PCM modulators results in a lower insertion loss and smaller chip area compared to using MZI modulators. For a circuit scale of $128 \times 128$, the estimated insertion loss is 56.9 dB for MZI modulators and 12.0 dB for PCM modulators, and the estimated chip area is 460 $\rm mm^2$ for MZI modulators and 86 $\rm mm^2$ for PCM modulators—both of which are smaller than the reticle size (858 $\rm mm^2$).

For a trained neural network, the weight matrices are fixed. In addition to their compact size, PCM modulators require no electrical power during static operations. Therefore, PCM modulators are preferable for fixed matrices. While the demonstrated circuit only performs linear operations, nonlinear activators can also be integrated to enable a multilayer optical neural network \cite{bandyopadhyay2022single}. For example, the photocurrent of the multiport PD can directly drive a p-i-n phase modulator in an MZI or MRR to realize electro-optic nonlinearity \cite{bandyopadhyay2022single, huang2022programmable}. The optical output of this MZI or MRR can then be used as an input for the next layer. By using compact and nonvolatile PCM modulators along with optical nonlinear activators, it is foreseeable that energy-efficient multilayer optical neural networks can be realized in the near future.  

\section{Conclusion}\label{sec4}
In conclusion, we have demonstrated a scalable photonic MVM processor based on the concept of waveguide multiplexing, which uses multiport PDs to sum the intensities of multiple optical signals. A 16-port Ge PD with a 3 dB bandwidth of 11.8 GHz at a bias voltage of -3 V was demonstrated, which can be further scaled up to handle 250 ports while maintaining a 6.1 GHz operation bandwidth. A $4 \times 4$ circuit fabricated on an SOI platform was used to perform MVMs in a 3-layer neural network designed for classifying Iris flowers, achieving a classification accuracy of 93.3\%. Furthermore, the performance of large-scale circuits in a CNN for Fashion-MNIST was simulated, resulting in a classification accuracy of 90.53\%. This work provides a simplified and scalable approach to photonic MVM, laying a foundation for large-scale and multi-dimensional photonic matrix-matrix multiplication in optical neural networks.

\section{Methods}\label{sec5}
\subsection{Design of multiport PD}
The design of a multiport PD involves a trade-off between bandwidth, dark current, and responsivity. The fabricated 16-port PD has a Ge area of $35 \times 13$ \unit{\micro\meter^2} and a 220-nm-thick Si terrace area of $39 \times 24.65$ \unit{\micro\meter^2}. Increasing the Ge length along the light propagation direction can enhance the responsivity by increasing light absorption, but it will also result in a higher dark current and lower bandwidth. The total width of the 16 input waveguides is 11.54 \unit{\micro\meter} (width: 440 nm, spacing: 300 nm), intentionally designed to be smaller than the Ge width (35 \unit{\micro\meter}) to ensure that light in waveguides far from the center is well absorbed. Two doping layers (p+ and p++) are used for Si in the multiport PD. The majority of the Si region is doped with the p+ layer, while the p++ layer is only used outside the Ge region.

\subsection{Characterization of multiport PD}
Light from a tunable laser (Santec, TSL 550) is amplified by an erbium-doped fiber amplifier (EDFA) (Luxpert, LXI 2000) and then injected into the 16-port PD via a grating coupler to characterize its responsivity and saturation current. The insertion loss of the grating coupler and the experimental setup has been excluded in the results shown in Fig. 2(e). We use a commercial optical intensity modulator (Sumicem, T.MXH1.5-20PD-ADC-LV) and a vector network analyzer (Keysight, N5225B) to characterize the bandwidth of the 16-port PD. The frequency response of a high-speed PD (3 dB bandwidth: $>$70 GHz) provided by AMF's process design kit (PDK) is first measured to calibrate the frequency response of the experimental setup.

\subsection{CNN for Fashion-MNIST}
The CNN shown in Fig. 6(a) consists of two convolutional layers and several fully connected layers. The rectified linear unit (ReLU) function is used as the nonlinear activation function in all layers. The parameters of the two convolutional layers are as follows: For layer 1, the kernel size is $3 \times 3$, the number of kernels is 30, the padding is 1, the stride is 1, and the pooling size is $2 \times 2$. For layer 2, the kernel size is $3 \times 3$, the number of kernels is 60, the padding is 0, the stride is 1, and the pooling size is $2 \times 2$. These two layers convert an input image with a size of $28 \times 28 \times 1$ into an output with a size of $6 \times 6 \times 60$, which is then flattened into a vector and processed by the fully-connected layers. PyTorch is used to construct this CNN, with cross entropy loss used as the loss function and Adam used as the optimizer. During the training, the learning rate is 0.001, the batch size is 100, and the number of epochs is 8. In addition, a dropout layer (drop probability: 0.25) is applied to the output of the 1st hidden layer of fully-connected layers to reduce overfitting during training. This dropout layer is not active during the inference phase. An Nvidia GPU (GeForce RTX 3080) is used to accelerate the training of this neural network. After training, the weight matrices in general include negative elements. For physical implementation, the weight matrices of the last two layers are linearly transformed into new matrices where all elements are normalized to values between 0 and 1, using a method described in our previous work \cite{ohno2022si}. 

\section*{Declarations}
\begin{itemize}
\item Funding: Japan Science and Technology Agency (CREST, JPMJCR2004); Japan Society for the Promotion of Science (22K14298).
\item Competing interests: The authors have reported preliminary results of this work at a conference \cite{tang2024single}.
\item Data availability: Data underlying the results presented in this paper are available from the corresponding authors upon reasonable request.
\item Code availability: Codes developed for Fashion-MNIST are available from the corresponding authors upon reasonable request.
\item Author contribution: R.T. conceived the idea, designed the circuit, performed the simulations and experiments, and wrote the manuscript. M.O. contributed to the device design, fabrication, and packaging. C.Z. assisted in characterizing the PD bandwidth. K.T. and S.T. contributed to discussions. M.T. revised the manuscript and supervised the project.
\end{itemize}

\begin{appendices}
\section{Light absorption in multiport PD}\label{secA}
\begin{figure}[t]
\centering
\includegraphics[width=0.9\textwidth]{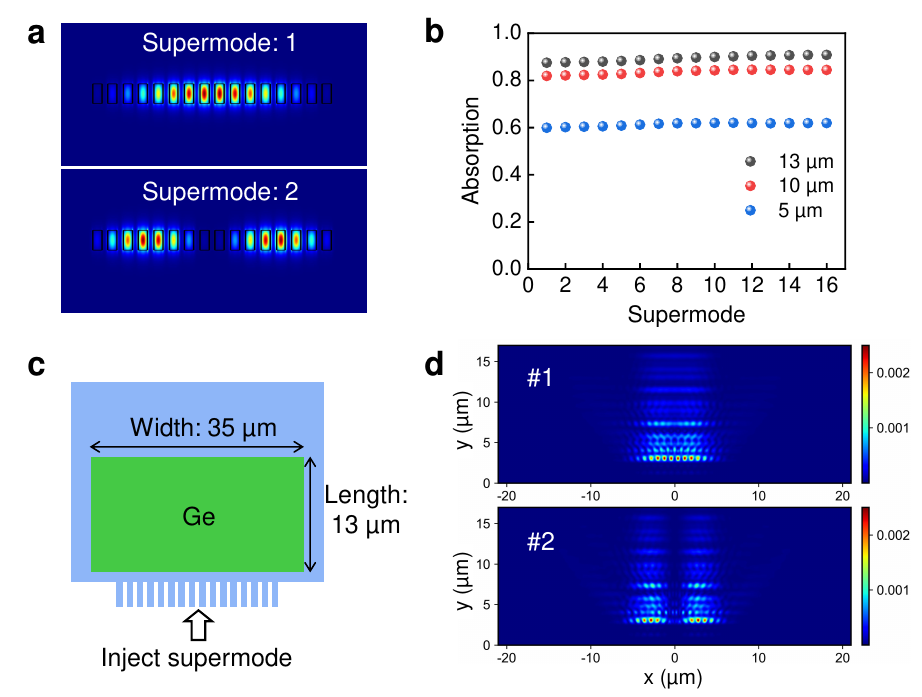}
\caption{\textbf{a} Profiles of the first two supermodes of a 16-waveguide array. \textbf{b} Simulated light absorption ratios for all supermodes in 16-port PDs with different Ge lengths. \textbf{c} Simulation of light propagation in the fabricated 16-port PD. \textbf{d} Simulated light propagation in the Ge region when the first and second supermodes are injected separately.}\label{fig-S1}
\end{figure}
Since the input waveguides of a multiport PD are adiabatically brought close to each other to form a multiport waveguide coupler (also known as coupled waveguide array), the incident light can be decomposed as a linear combination of the supermodes of the waveguide array. The number of supermodes per polarization is exactly the same as the number of waveguides. Figure A1(a) shows the profiles of the first two supermodes of a 16-waveguide array, which is used in the 16-port PD. We further simulate the light absorption in 16-port PDs with different Ge lengths when each of the 16 supermodes is injected individually. Figure A1(b) shows the results for Ge lengths of 5, 10, and 13 µm, with a Ge width of 35 µm. The Ge length in our fabricated 16-port PD is 13 µm. It can be seen that all supermodes have similar absorption ratios for a given length. Slight differences are mainly attributed to the small variations in the effective indices of these supermodes. We then simulate the light propagation in the fabricated 16-port PD, as illustrated in Fig. A1(c). Here, to simplify the simulations, the doped Si, doped Ge, and electrode regions are not considered. Therefore, light attenuation should be stronger in real devices than in this simulation. Figure A1(d) shows the simulated light propagation in the Ge region when the first and second supermodes are injected separately. As expected, the light gradually attenuates along its propagation. The intensity distributions are significantly different for these two supermodes, but the absorption ratios are very close.

\section{Dark current of multiport PD}\label{secB}
\begin{figure}[b]
\centering
\includegraphics[width=1.0\textwidth]{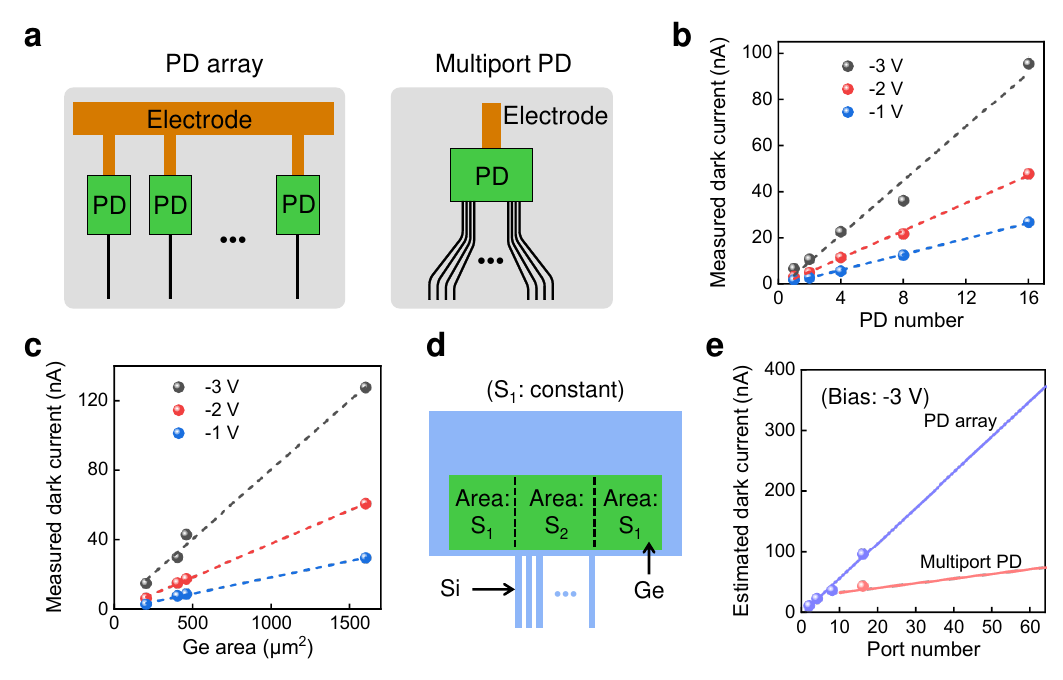}
\caption{\textbf{a} Adding up multiple optical signals with a single-port PD array and a multiport PD, respectively. \textbf{b} Measured dark currents of single-port PD arrays. \textbf{c} Measured dark currents of custom-designed PDs with Ge/Si structures similar to that of the 16-port PD. \textbf{d} Estimating the dark current of multiport PD from the Ge area. \textbf{e} Estimated dark currents of single-port PD arrays and multiport PDs.}\label{fig-S2}
\end{figure}
For Ge PDs using the same structure, the dark current is generally proportional to the area of the Ge region. In this scheme, the addition of multiple optical signals can also be achieved using a single-port PD array, as illustrated in Fig. B2(a). Each PD detects a single optical signal, and the photocurrents from multiple PDs are combined by a common electrode. Compared with a large-scale PD array, a multiport PD with the same number of input ports can have a smaller Ge area, resulting in a smaller dark current and capacitance. Figure B2(b) shows the measured dark currents of single-port PD arrays of various scales, using the PD provided by the process design kit (PDK) of AMF \cite{siew2021review}. As expected, the total dark current is approximately the dark current of a single PD multiplied by the number of PDs. For the 16-PD array, the measured dark current is 95.8 nA at a -3 V bias voltage. Figure B2(c) shows the measured dark currents of our custom-designed PDs, which have Ge/Si structures similar to that of the 16-port PD. The dark currents are observed to be approximately proportional to the Ge area. The measured dark current of the 16-port PD is 43 nA at a -3 V bias voltage, which is approximately half that of the 16-PD array. We further estimate the dark current of multiport PDs using a simple model depicted in Fig. B2(d). We assume that the area of the Ge regions on the left and right sides ($S_1$ in the figure) are constant and use the same value as that of the 16-port PD. The area of the middle Ge region ($S_2$ in the figure) is determined by the number of waveguides and their pitch. We also assume the same waveguide pitch (740 nm) as the 16-port PD, consisting of a 440-nm waveguide width and 300-nm spacing. The dark current is then estimated from the Ge area, using the slope extracted from Fig. B2(c). The result is shown in Fig. B2(e). It can be seen that the increase in dark current is significantly slower in multiport PDs compared to the PD array as the number of input ports increases.

\section{Performance of a large Ge PD}\label{secC}
A large Ge PD is designed and fabricated using a structure similar to the multiport PD, as shown in Fig. C3(a-b). The Ge area is $200 \times 8$ \unit{\micro\meter^2}. The width of the input waveguide is linearly increased to match the Ge width using a 400-µm-long taper. Figure C3(c) shows the measured dark current as a function of the bias voltage. The dark current is 127.7 nA at a -3 V bias voltage. Figure C3(d) shows the measured frequency response of the PD, exhibiting a 3 dB bandwidth of approximately 6.1 GHz at a -3 V bias voltage. Note that the dark current and frequency response do not depend on the number of input ports if the Ge area and other parameters remain the same. Although this PD only has one input port, the large Ge width can approximately support 250 input ports, given a waveguide pitch of 740 nm.

\begin{figure}[H]
\centering
\includegraphics[width=0.8\textwidth]{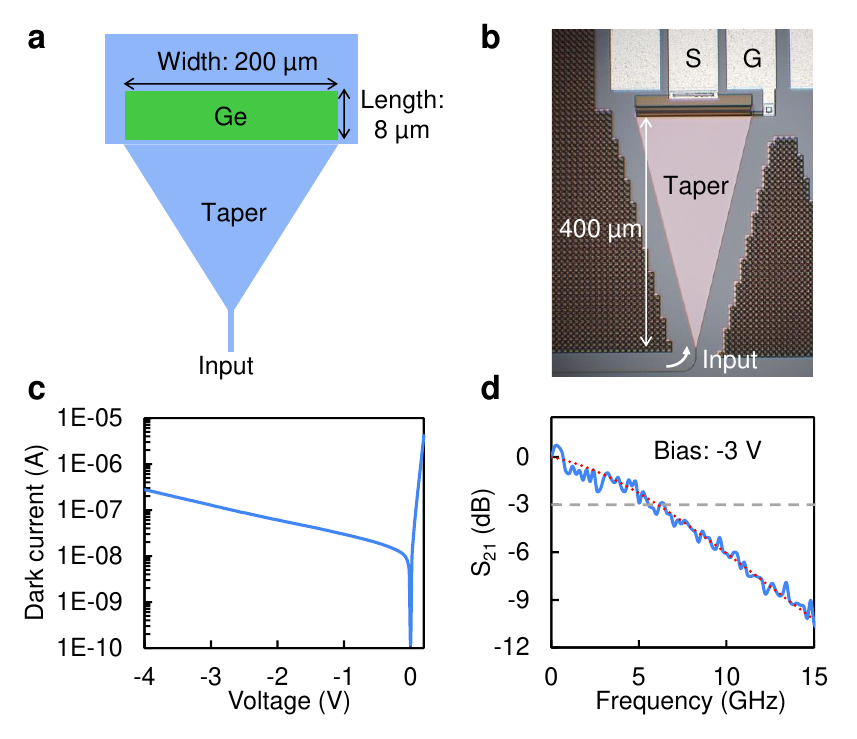}
\caption{\textbf{a} Schematic structure (top view) of a large Ge PD. The drawing is not to scale. Doped regions and electrodes are not shown. \textbf{b} A microscope image of the fabricated PD. \textbf{c} Measured dark current as a function of the bias voltage. \textbf{d} Measured electro-optic frequency response of the PD at a -3 V bias voltage, showing a 3 dB bandwidth of approximately 6.1 GHz.}\label{fig-S3}
\end{figure}

\section{Compatibility with wavelength and mode multiplexing}\label{secD}
The compatibility of this scheme with wavelength multiplexing was proposed and discussed in our original proposal \cite{tang2022two}. Here, we provide a schematic structure in Fig. D4 to illustrate its compatibility with mode multiplexing. Multiple vectors (two vectors in Fig. D4) are generated and then multiplexed into multimode waveguides using adiabatic mode (de)multiplexers \cite{ling2023chip}. These multimode waveguides are equally split, and the light within them are further modulated by the intensity modulators for the matrix. These modulators should provide equal intensity modulation for all modes used in the multimode waveguide \cite{mojaver2024recent}. After this, the light encoded on multiple waveguide modes are demultiplexed and then detected by multiport PDs, performing matrix-matrix multiplication (parallel MVMs). Compared to the previous work using mode multiplexing \cite{ling2023chip}, the use of mode multiplexing in this scheme does not significantly increase the insertion loss.

\begin{figure}[H]
\centering
\includegraphics[width=1\textwidth]{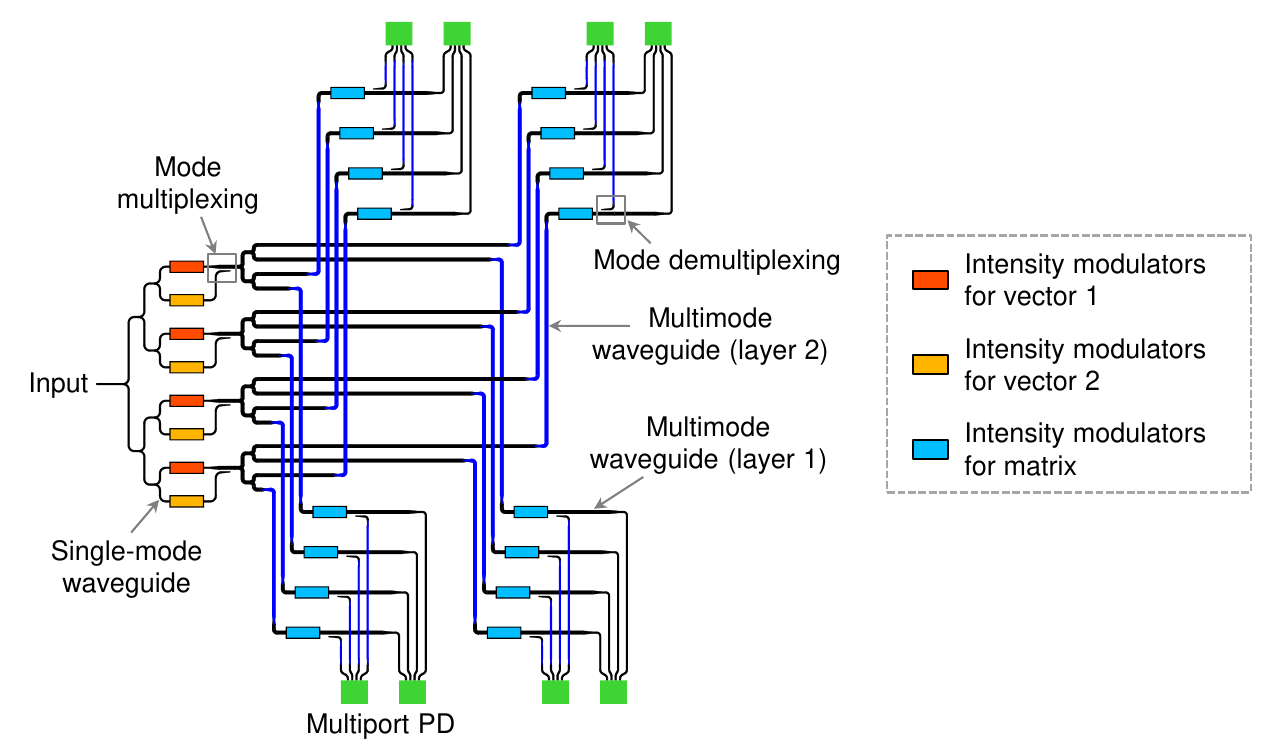}
\caption{Adding the waveguide mode dimension to this scheme enables matrix-matrix multiplication (parallel MVMs). The intensity modulators for the matrix should provide equal intensity modulation for all modes used in the multimode waveguide.}\label{fig-S4}
\end{figure}

\section{Estimated chip area and insertion loss}\label{secE}
We estimate the chip area and insertion loss for the structure shown in Fig. E5(a). Here, to reduce the number of waveguides, light is split using directional couplers instead of multi-stage optical splitters. Intensity modulators can vary in size and insertion loss. In this calculation, MZIs and PCM absorbers are assumed as the intensity modulators separately. Advanced packaging techniques, such as through-silicon via (TSV), can eliminate the need for complicated metal electrodes, so only the intrinsic area of optical components (modulators, waveguides, and PDs) is considered. In addition, the loss includes matrix loss and extra loss caused by optical components. Since the matrix loss is independent of chip structures, only the extra loss is considered. The parameters used in the calculation are shown in Table E1.

The chip area is calculated according to
\begin{equation}
\begin{split}
{\rm S} &= {\rm S_{mod} + S_{wg} + S_{PD}}\\
&= (N^2+N){\rm S_{MZI/PCM}} + N{\rm L_{1}} \cdot N{\rm L_{3}}/2 + N{\rm L_{3}(L_{4} + L_{5})},
\end{split}
\end{equation}
where $\rm S_{mod}$, $\rm S_{wg}$, and $\rm S_{PD}$ represent the total area of intensity modulators, waveguides, and multiport PDs, respectively, $N$ is the circuit or matrix scale. The insertion loss for the longest path is calculated according to
\begin{equation}
\begin{split}
{\rm IL} &= {\rm IL_{mod} + IL_{wg} + IL_{other}}\\
&= 2{\rm IL_{MZI/PCM}} + (N^2{\rm L_{2}}/2 + N{\rm L_{1}} + \sqrt{2}{\rm L_{3}})\alpha + {\rm IL_{other}},
\end{split}
\end{equation}
where $\rm IL_{mod}$, $\rm IL_{wg}$, and $\rm IL_{other}$ represent the static loss of intensity modulators, light propagation loss, and extra loss of other components, respectively, $\alpha$ is the waveguide propagation loss per unit length. Figure E5(b) and E5(b) show the estimated chip area and insertion loss, respectively. Since a PCM modulator is significantly more compact and has a lower static loss than an MZI modulator, implementing this scheme using PCM modulators leads to a lower insertion loss and smaller chip area than using MZI modulators. Assuming a circuit scale of $128 \times 128$, the estimated insertion loss is 56.9 dB for MZI modulators and 12.0 dB for PCM modulators, the estimated chip area is 460 $\rm mm^2$ for MZI modulators and 86 $\rm mm^2$ for PCM modulators, which are smaller than the reticle size (858 $\rm mm^2$).

\begin{table}[h]
\caption{Parameters used to estimate the chip area and insertion loss}\label{tab2}%
\begin{tabular}{@{}lll@{}}
\toprule
Parameters & Value & Note \\
\midrule
$\rm{L_1}$ & 8 µm &  \\
$\rm{S_{MZI}}$ & $300 \times 50$ \unit{\micro\meter^2} & MZI modulator area \\
$\rm{S_{PCM}}$ & $40 \times 10$ \unit{\micro\meter^2} & PCM modulator area \\
$\rm{L_2}$ (MZI) & 50 µm &  \\
$\rm{L_2}$ (PCM) & 10 µm &  \\
$\rm{L_3}$ & $(N-1)\rm{L_2}$ &  \\
$\rm{L_4}$ & $\rm{L_3}$ &  \\
$\rm{L_5}$ & 50 µm & Waveguide length not included \\
$\alpha$ & 1.3 dB/cm & Waveguide propagation loss \\
$\rm{IL_{MZI}}$ & 1 dB & Static loss of MZI modulators \\
$\rm{IL_{PCM}}$ & 0.3 dB & Static loss of PCM modulators \\
$\rm{IL_{other}}$ & 0.4 dB & \makecell[c]{0.1 dB (directional coupler) \\ 0.1 dB $\times$ 2 (interlayer coupler) \\ 0.1 dB (optical splitter)} \\
\botrule
\end{tabular}
\end{table}

\begin{figure}[H]
\centering
\includegraphics[width=1\textwidth]{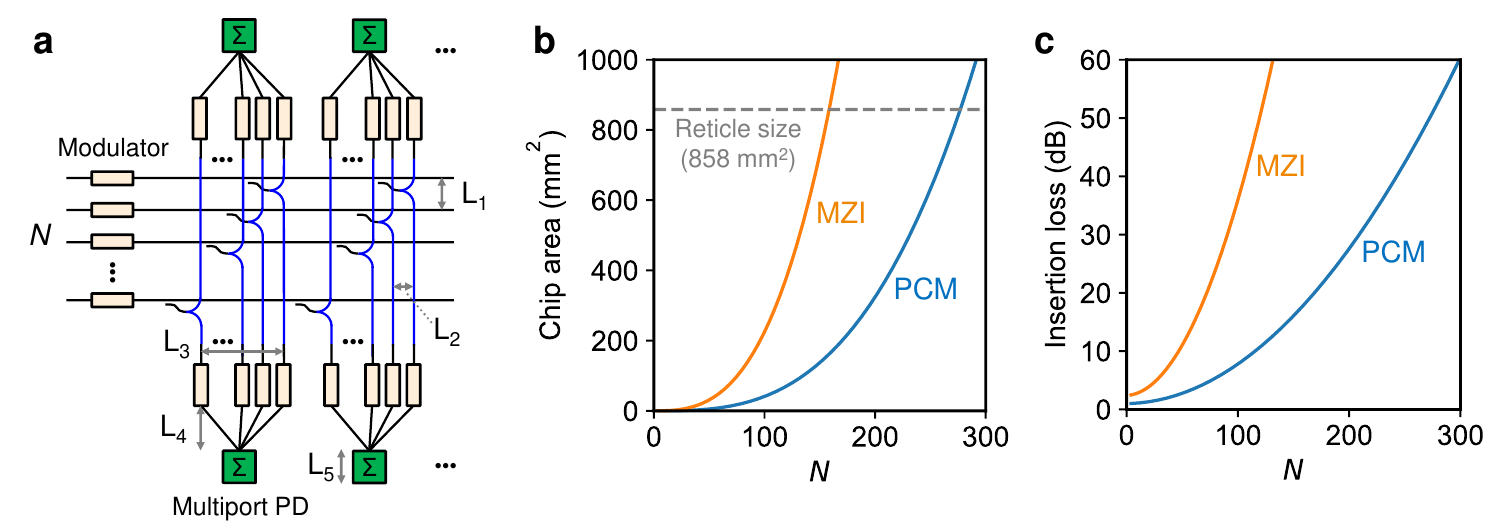}
\caption{\textbf{a} Schematic structure for large-scale circuits. Directional couplers and multiple waveguide layers are used to reduce the circuit size. \textbf{b} Estimated chip area as a function of circuit scale, assuming MZIs and PCM absorbers are used as the intensity modulators separately. \textbf{c} Estimated insertion loss as a function of circuit scale, assuming MZIs and PCM absorbers are used as the intensity modulators separately.}\label{fig-S5}
\end{figure}




\end{appendices}


\bibliography{sn-bibliography}


\begin{thebibliography}{50}
\ifx \bisbn   \undefined \def \bisbn  #1{ISBN #1}\fi
\ifx \binits  \undefined \def \binits#1{#1}\fi
\ifx \bauthor  \undefined \def \bauthor#1{#1}\fi
\ifx \batitle  \undefined \def \batitle#1{#1}\fi
\ifx \bjtitle  \undefined \def \bjtitle#1{#1}\fi
\ifx \bvolume  \undefined \def \bvolume#1{\textbf{#1}}\fi
\ifx \byear  \undefined \def \byear#1{#1}\fi
\ifx \bissue  \undefined \def \bissue#1{#1}\fi
\ifx \bfpage  \undefined \def \bfpage#1{#1}\fi
\ifx \blpage  \undefined \def \blpage #1{#1}\fi
\ifx \burl  \undefined \def \burl#1{\textsf{#1}}\fi
\ifx \doiurl  \undefined \def \doiurl#1{\url{https://doi.org/#1}}\fi
\ifx \betal  \undefined \def \betal{\textit{et al.}}\fi
\ifx \binstitute  \undefined \def \binstitute#1{#1}\fi
\ifx \binstitutionaled  \undefined \def \binstitutionaled#1{#1}\fi
\ifx \bctitle  \undefined \def \bctitle#1{#1}\fi
\ifx \beditor  \undefined \def \beditor#1{#1}\fi
\ifx \bpublisher  \undefined \def \bpublisher#1{#1}\fi
\ifx \bbtitle  \undefined \def \bbtitle#1{#1}\fi
\ifx \bedition  \undefined \def \bedition#1{#1}\fi
\ifx \bseriesno  \undefined \def \bseriesno#1{#1}\fi
\ifx \blocation  \undefined \def \blocation#1{#1}\fi
\ifx \bsertitle  \undefined \def \bsertitle#1{#1}\fi
\ifx \bsnm \undefined \def \bsnm#1{#1}\fi
\ifx \bsuffix \undefined \def \bsuffix#1{#1}\fi
\ifx \bparticle \undefined \def \bparticle#1{#1}\fi
\ifx \barticle \undefined \def \barticle#1{#1}\fi
\bibcommenthead
\ifx \bconfdate \undefined \def \bconfdate #1{#1}\fi
\ifx \botherref \undefined \def \botherref #1{#1}\fi
\ifx \url \undefined \def \url#1{\textsf{#1}}\fi
\ifx \bchapter \undefined \def \bchapter#1{#1}\fi
\ifx \bbook \undefined \def \bbook#1{#1}\fi
\ifx \bcomment \undefined \def \bcomment#1{#1}\fi
\ifx \oauthor \undefined \def \oauthor#1{#1}\fi
\ifx \citeauthoryear \undefined \def \citeauthoryear#1{#1}\fi
\ifx \endbibitem  \undefined \def \endbibitem {}\fi
\ifx \bconflocation  \undefined \def \bconflocation#1{#1}\fi
\ifx \arxivurl  \undefined \def \arxivurl#1{\textsf{#1}}\fi
\csname PreBibitemsHook\endcsname

\bibitem[\protect\citeauthoryear{Shen et~al.}{2017}]{shen2017deep}
\begin{barticle}
\bauthor{\bsnm{Shen}, \binits{Y.}},
\bauthor{\bsnm{Harris}, \binits{N.C.}},
\bauthor{\bsnm{Skirlo}, \binits{S.}},
\bauthor{\bsnm{Prabhu}, \binits{M.}},
\bauthor{\bsnm{Baehr-Jones}, \binits{T.}},
\bauthor{\bsnm{Hochberg}, \binits{M.}},
\bauthor{\bsnm{Sun}, \binits{X.}},
\bauthor{\bsnm{Zhao}, \binits{S.}},
\bauthor{\bsnm{Larochelle}, \binits{H.}},
\bauthor{\bsnm{Englund}, \binits{D.}},
\bauthor{\bsnm{Soljačić}, \binits{M.}}:
\batitle{Deep learning with coherent nanophotonic circuits}.
\bjtitle{Nat. Photon.}
\bvolume{11}(\bissue{7}),
\bfpage{441}--\blpage{446}
(\byear{2017})
\end{barticle}
\endbibitem

\bibitem[\protect\citeauthoryear{Lin et~al.}{2018}]{lin2018all}
\begin{barticle}
\bauthor{\bsnm{Lin}, \binits{X.}},
\bauthor{\bsnm{Rivenson}, \binits{Y.}},
\bauthor{\bsnm{Yardimci}, \binits{N.T.}},
\bauthor{\bsnm{Veli}, \binits{M.}},
\bauthor{\bsnm{Luo}, \binits{Y.}},
\bauthor{\bsnm{Jarrahi}, \binits{M.}},
\bauthor{\bsnm{Ozcan}, \binits{A.}}:
\batitle{All-optical machine learning using diffractive deep neural networks}.
\bjtitle{Science}
\bvolume{361}(\bissue{6406}),
\bfpage{1004}--\blpage{1008}
(\byear{2018})
\end{barticle}
\endbibitem

\bibitem[\protect\citeauthoryear{Hamerly et~al.}{2019}]{hamerly2019large}
\begin{barticle}
\bauthor{\bsnm{Hamerly}, \binits{R.}},
\bauthor{\bsnm{Bernstein}, \binits{L.}},
\bauthor{\bsnm{Sludds}, \binits{A.}},
\bauthor{\bsnm{Solja{\v{c}}i{\'c}}, \binits{M.}},
\bauthor{\bsnm{Englund}, \binits{D.}}:
\batitle{Large-scale optical neural networks based on photoelectric multiplication}.
\bjtitle{Phys. Rev. X}
\bvolume{9}(\bissue{2}),
\bfpage{021032}
(\byear{2019})
\end{barticle}
\endbibitem

\bibitem[\protect\citeauthoryear{Wetzstein et~al.}{2020}]{wetzstein2020inference}
\begin{barticle}
\bauthor{\bsnm{Wetzstein}, \binits{G.}},
\bauthor{\bsnm{Ozcan}, \binits{A.}},
\bauthor{\bsnm{Gigan}, \binits{S.}},
\bauthor{\bsnm{Fan}, \binits{S.}},
\bauthor{\bsnm{Englund}, \binits{D.}},
\bauthor{\bsnm{Solja{\v{c}}i{\'c}}, \binits{M.}},
\bauthor{\bsnm{Denz}, \binits{C.}},
\bauthor{\bsnm{Miller}, \binits{D.A.}},
\bauthor{\bsnm{Psaltis}, \binits{D.}}:
\batitle{Inference in artificial intelligence with deep optics and photonics}.
\bjtitle{Nature}
\bvolume{588}(\bissue{7836}),
\bfpage{39}--\blpage{47}
(\byear{2020})
\end{barticle}
\endbibitem

\bibitem[\protect\citeauthoryear{Shastri et~al.}{2021}]{shastri2021photonics}
\begin{barticle}
\bauthor{\bsnm{Shastri}, \binits{B.J.}},
\bauthor{\bsnm{Tait}, \binits{A.N.}},
\bauthor{\bsnm{Lima}, \binits{T.}},
\bauthor{\bsnm{Pernice}, \binits{W.H.}},
\bauthor{\bsnm{Bhaskaran}, \binits{H.}},
\bauthor{\bsnm{Wright}, \binits{C.D.}},
\bauthor{\bsnm{Prucnal}, \binits{P.R.}}:
\batitle{Photonics for artificial intelligence and neuromorphic computing}.
\bjtitle{Nat. Photon.}
\bvolume{15}(\bissue{2}),
\bfpage{102}--\blpage{114}
(\byear{2021})
\end{barticle}
\endbibitem

\bibitem[\protect\citeauthoryear{Cong et~al.}{2022}]{cong2022chip}
\begin{barticle}
\bauthor{\bsnm{Cong}, \binits{G.}},
\bauthor{\bsnm{Yamamoto}, \binits{N.}},
\bauthor{\bsnm{Inoue}, \binits{T.}},
\bauthor{\bsnm{Maegami}, \binits{Y.}},
\bauthor{\bsnm{Ohno}, \binits{M.}},
\bauthor{\bsnm{Kita}, \binits{S.}},
\bauthor{\bsnm{Namiki}, \binits{S.}},
\bauthor{\bsnm{Yamada}, \binits{K.}}:
\batitle{On-chip bacterial foraging training in silicon photonic circuits for projection-enabled nonlinear classification}.
\bjtitle{Nat. Commun.}
\bvolume{13}(\bissue{1}),
\bfpage{3261}
(\byear{2022})
\end{barticle}
\endbibitem

\bibitem[\protect\citeauthoryear{Zhou et~al.}{2022}]{zhou2022photonic}
\begin{barticle}
\bauthor{\bsnm{Zhou}, \binits{H.}},
\bauthor{\bsnm{Dong}, \binits{J.}},
\bauthor{\bsnm{Cheng}, \binits{J.}},
\bauthor{\bsnm{Dong}, \binits{W.}},
\bauthor{\bsnm{Huang}, \binits{C.}},
\bauthor{\bsnm{Shen}, \binits{Y.}},
\bauthor{\bsnm{Zhang}, \binits{Q.}},
\bauthor{\bsnm{Gu}, \binits{M.}},
\bauthor{\bsnm{Qian}, \binits{C.}},
\bauthor{\bsnm{Chen}, \binits{H.}},
\bauthor{\bsnm{Ruan}, \binits{Z.}},
\bauthor{\bsnm{Zhang}, \binits{X.}}:
\batitle{Photonic matrix multiplication lights up photonic accelerator and beyond}.
\bjtitle{Light Sci. Appl.}
\bvolume{11}(\bissue{1}),
\bfpage{30}
(\byear{2022})
\end{barticle}
\endbibitem

\bibitem[\protect\citeauthoryear{Wang et~al.}{2022}]{wang2022optical}
\begin{barticle}
\bauthor{\bsnm{Wang}, \binits{T.}},
\bauthor{\bsnm{Ma}, \binits{S.-Y.}},
\bauthor{\bsnm{Wright}, \binits{L.G.}},
\bauthor{\bsnm{Onodera}, \binits{T.}},
\bauthor{\bsnm{Richard}, \binits{B.C.}},
\bauthor{\bsnm{McMahon}, \binits{P.L.}}:
\batitle{An optical neural network using less than 1 photon per multiplication}.
\bjtitle{Nat. Commun.}
\bvolume{13}(\bissue{1}),
\bfpage{123}
(\byear{2022})
\end{barticle}
\endbibitem

\bibitem[\protect\citeauthoryear{Ashtiani et~al.}{2022}]{ashtiani2022chip}
\begin{barticle}
\bauthor{\bsnm{Ashtiani}, \binits{F.}},
\bauthor{\bsnm{Geers}, \binits{A.J.}},
\bauthor{\bsnm{Aflatouni}, \binits{F.}}:
\batitle{An on-chip photonic deep neural network for image classification}.
\bjtitle{Nature}
\bvolume{606}(\bissue{7914}),
\bfpage{501}--\blpage{506}
(\byear{2022})
\end{barticle}
\endbibitem

\bibitem[\protect\citeauthoryear{Zhu et~al.}{2022}]{zhu2022space}
\begin{barticle}
\bauthor{\bsnm{Zhu}, \binits{H.}},
\bauthor{\bsnm{Zou}, \binits{J.}},
\bauthor{\bsnm{Zhang}, \binits{H.}},
\bauthor{\bsnm{Shi}, \binits{Y.}},
\bauthor{\bsnm{Luo}, \binits{S.}},
\bauthor{\bsnm{Wang}, \binits{N.}},
\bauthor{\bsnm{Cai}, \binits{H.}},
\bauthor{\bsnm{Wan}, \binits{L.}},
\bauthor{\bsnm{Wang}, \binits{B.}},
\bauthor{\bsnm{Jiang}, \binits{X.}}, \betal:
\batitle{Space-efficient optical computing with an integrated chip diffractive neural network}.
\bjtitle{Nat. Commun.}
\bvolume{13}(\bissue{1}),
\bfpage{1044}
(\byear{2022})
\end{barticle}
\endbibitem

\bibitem[\protect\citeauthoryear{Bernstein et~al.}{2023}]{bernstein2023single}
\begin{barticle}
\bauthor{\bsnm{Bernstein}, \binits{L.}},
\bauthor{\bsnm{Sludds}, \binits{A.}},
\bauthor{\bsnm{Panuski}, \binits{C.}},
\bauthor{\bsnm{Trajtenberg-Mills}, \binits{S.}},
\bauthor{\bsnm{Hamerly}, \binits{R.}},
\bauthor{\bsnm{Englund}, \binits{D.}}:
\batitle{Single-shot optical neural network}.
\bjtitle{Sci. Adv.}
\bvolume{9}(\bissue{25}),
\bfpage{7904}
(\byear{2023})
\end{barticle}
\endbibitem

\bibitem[\protect\citeauthoryear{McMahon}{2023}]{mcmahon2023physics}
\begin{barticle}
\bauthor{\bsnm{McMahon}, \binits{P.L.}}:
\batitle{The physics of optical computing}.
\bjtitle{Nat. Rev. Phys.}
\bvolume{5}(\bissue{12}),
\bfpage{717}--\blpage{734}
(\byear{2023})
\end{barticle}
\endbibitem

\bibitem[\protect\citeauthoryear{Chen et~al.}{2023a}]{chen2023deep}
\begin{barticle}
\bauthor{\bsnm{Chen}, \binits{Z.}},
\bauthor{\bsnm{Sludds}, \binits{A.}},
\bauthor{\bsnm{Davis~III}, \binits{R.}},
\bauthor{\bsnm{Christen}, \binits{I.}},
\bauthor{\bsnm{Bernstein}, \binits{L.}},
\bauthor{\bsnm{Ateshian}, \binits{L.}},
\bauthor{\bsnm{Heuser}, \binits{T.}},
\bauthor{\bsnm{Heermeier}, \binits{N.}},
\bauthor{\bsnm{Lott}, \binits{J.A.}},
\bauthor{\bsnm{Reitzenstein}, \binits{S.}}, \betal:
\batitle{Deep learning with coherent {VCSEL} neural networks}.
\bjtitle{Nat. Photon.}
\bvolume{17}(\bissue{8}),
\bfpage{723}--\blpage{730}
(\byear{2023})
\end{barticle}
\endbibitem

\bibitem[\protect\citeauthoryear{Chen et~al.}{2023b}]{chen2023all}
\begin{barticle}
\bauthor{\bsnm{Chen}, \binits{Y.}},
\bauthor{\bsnm{Nazhamaiti}, \binits{M.}},
\bauthor{\bsnm{Xu}, \binits{H.}},
\bauthor{\bsnm{Meng}, \binits{Y.}},
\bauthor{\bsnm{Zhou}, \binits{T.}},
\bauthor{\bsnm{Li}, \binits{G.}},
\bauthor{\bsnm{Fan}, \binits{J.}},
\bauthor{\bsnm{Wei}, \binits{Q.}},
\bauthor{\bsnm{Wu}, \binits{J.}},
\bauthor{\bsnm{Qiao}, \binits{F.}}, \betal:
\batitle{All-analog photoelectronic chip for high-speed vision tasks}.
\bjtitle{Nature}
\bvolume{623}(\bissue{7985}),
\bfpage{48}--\blpage{57}
(\byear{2023})
\end{barticle}
\endbibitem

\bibitem[\protect\citeauthoryear{Xu et~al.}{2024}]{xu2024large}
\begin{barticle}
\bauthor{\bsnm{Xu}, \binits{Z.}},
\bauthor{\bsnm{Zhou}, \binits{T.}},
\bauthor{\bsnm{Ma}, \binits{M.}},
\bauthor{\bsnm{Deng}, \binits{C.}},
\bauthor{\bsnm{Dai}, \binits{Q.}},
\bauthor{\bsnm{Fang}, \binits{L.}}:
\batitle{Large-scale photonic chiplet {Taichi} empowers 160-{TOPS/W} artificial general intelligence}.
\bjtitle{Science}
\bvolume{384}(\bissue{6692}),
\bfpage{202}--\blpage{209}
(\byear{2024})
\end{barticle}
\endbibitem

\bibitem[\protect\citeauthoryear{Xue et~al.}{2024}]{xue2024fully}
\begin{barticle}
\bauthor{\bsnm{Xue}, \binits{Z.}},
\bauthor{\bsnm{Zhou}, \binits{T.}},
\bauthor{\bsnm{Xu}, \binits{Z.}},
\bauthor{\bsnm{Yu}, \binits{S.}},
\bauthor{\bsnm{Dai}, \binits{Q.}},
\bauthor{\bsnm{Fang}, \binits{L.}}:
\batitle{Fully forward mode training for optical neural networks}.
\bjtitle{Nature}
\bvolume{632}(\bissue{8024}),
\bfpage{280}--\blpage{286}
(\byear{2024})
\end{barticle}
\endbibitem

\bibitem[\protect\citeauthoryear{Cheng et~al.}{2024}]{cheng2024multimodal}
\begin{barticle}
\bauthor{\bsnm{Cheng}, \binits{J.}},
\bauthor{\bsnm{Huang}, \binits{C.}},
\bauthor{\bsnm{Zhang}, \binits{J.}},
\bauthor{\bsnm{Wu}, \binits{B.}},
\bauthor{\bsnm{Zhang}, \binits{W.}},
\bauthor{\bsnm{Liu}, \binits{X.}},
\bauthor{\bsnm{Zhang}, \binits{J.}},
\bauthor{\bsnm{Tang}, \binits{Y.}},
\bauthor{\bsnm{Zhou}, \binits{H.}},
\bauthor{\bsnm{Zhang}, \binits{Q.}}, \betal:
\batitle{Multimodal deep learning using on-chip diffractive optics with in situ training capability}.
\bjtitle{Nat. Commun.}
\bvolume{15}(\bissue{1}),
\bfpage{6189}
(\byear{2024})
\end{barticle}
\endbibitem

\bibitem[\protect\citeauthoryear{Clements et~al.}{2016}]{clements2016optimal}
\begin{barticle}
\bauthor{\bsnm{Clements}, \binits{W.R.}},
\bauthor{\bsnm{Humphreys}, \binits{P.C.}},
\bauthor{\bsnm{Metcalf}, \binits{B.J.}},
\bauthor{\bsnm{Kolthammer}, \binits{W.S.}},
\bauthor{\bsnm{Walmsley}, \binits{I.A.}}:
\batitle{Optimal design for universal multiport interferometers}.
\bjtitle{Optica}
\bvolume{3}(\bissue{12}),
\bfpage{1460}--\blpage{1465}
(\byear{2016})
\end{barticle}
\endbibitem

\bibitem[\protect\citeauthoryear{Tang et~al.}{2021}]{tang2021ten}
\begin{barticle}
\bauthor{\bsnm{Tang}, \binits{R.}},
\bauthor{\bsnm{Tanomura}, \binits{R.}},
\bauthor{\bsnm{Tanemura}, \binits{T.}},
\bauthor{\bsnm{Nakano}, \binits{Y.}}:
\batitle{Ten-port unitary optical processor on a silicon photonic chip}.
\bjtitle{ACS Photonics}
\bvolume{8}(\bissue{7}),
\bfpage{2074}--\blpage{2080}
(\byear{2021})
\end{barticle}
\endbibitem

\bibitem[\protect\citeauthoryear{Zhang et~al.}{2021}]{zhang2021optical}
\begin{barticle}
\bauthor{\bsnm{Zhang}, \binits{H.}},
\bauthor{\bsnm{Gu}, \binits{M.}},
\bauthor{\bsnm{Jiang}, \binits{X.}},
\bauthor{\bsnm{Thompson}, \binits{J.}},
\bauthor{\bsnm{Cai}, \binits{H.}},
\bauthor{\bsnm{Paesani}, \binits{S.}},
\bauthor{\bsnm{Santagati}, \binits{R.}},
\bauthor{\bsnm{Laing}, \binits{A.}},
\bauthor{\bsnm{Zhang}, \binits{Y.}},
\bauthor{\bsnm{Yung}, \binits{M.-H.}}, \betal:
\batitle{An optical neural chip for implementing complex-valued neural network}.
\bjtitle{Nat. Commun.}
\bvolume{12}(\bissue{1}),
\bfpage{457}
(\byear{2021})
\end{barticle}
\endbibitem

\bibitem[\protect\citeauthoryear{Hamerly et~al.}{2022}]{hamerly2022asymptotically}
\begin{barticle}
\bauthor{\bsnm{Hamerly}, \binits{R.}},
\bauthor{\bsnm{Bandyopadhyay}, \binits{S.}},
\bauthor{\bsnm{Englund}, \binits{D.}}:
\batitle{Asymptotically fault-tolerant programmable photonics}.
\bjtitle{Nat. Commun.}
\bvolume{13}(\bissue{1}),
\bfpage{6831}
(\byear{2022})
\end{barticle}
\endbibitem

\bibitem[\protect\citeauthoryear{Mourgias-Alexandris et~al.}{2022}]{mourgias2022noise}
\begin{barticle}
\bauthor{\bsnm{Mourgias-Alexandris}, \binits{G.}},
\bauthor{\bsnm{Moralis-Pegios}, \binits{M.}},
\bauthor{\bsnm{Tsakyridis}, \binits{A.}},
\bauthor{\bsnm{Simos}, \binits{S.}},
\bauthor{\bsnm{Dabos}, \binits{G.}},
\bauthor{\bsnm{Totovic}, \binits{A.}},
\bauthor{\bsnm{Passalis}, \binits{N.}},
\bauthor{\bsnm{Kirtas}, \binits{M.}},
\bauthor{\bsnm{Rutirawut}, \binits{T.}},
\bauthor{\bsnm{Gardes}, \binits{F.}}, \betal:
\batitle{Noise-resilient and high-speed deep learning with coherent silicon photonics}.
\bjtitle{Nat. Commun.}
\bvolume{13}(\bissue{1}),
\bfpage{5572}
(\byear{2022})
\end{barticle}
\endbibitem

\bibitem[\protect\citeauthoryear{Bandyopadhyay et~al.}{2022}]{bandyopadhyay2022single}
\begin{botherref}
\oauthor{\bsnm{Bandyopadhyay}, \binits{S.}},
\oauthor{\bsnm{Sludds}, \binits{A.}},
\oauthor{\bsnm{Krastanov}, \binits{S.}},
\oauthor{\bsnm{Hamerly}, \binits{R.}},
\oauthor{\bsnm{Harris}, \binits{N.}},
\oauthor{\bsnm{Bunandar}, \binits{D.}},
\oauthor{\bsnm{Streshinsky}, \binits{M.}},
\oauthor{\bsnm{Hochberg}, \binits{M.}},
\oauthor{\bsnm{Englund}, \binits{D.}}:
Single chip photonic deep neural network with accelerated training.
arXiv:2208.01623
(2022)
\end{botherref}
\endbibitem

\bibitem[\protect\citeauthoryear{Giamougiannis et~al.}{2023}]{giamougiannis2023coherent}
\begin{barticle}
\bauthor{\bsnm{Giamougiannis}, \binits{G.}},
\bauthor{\bsnm{Tsakyridis}, \binits{A.}},
\bauthor{\bsnm{Ma}, \binits{Y.}},
\bauthor{\bsnm{Totovi{\'c}}, \binits{A.}},
\bauthor{\bsnm{Moralis-Pegios}, \binits{M.}},
\bauthor{\bsnm{Lazovsky}, \binits{D.}},
\bauthor{\bsnm{Pleros}, \binits{N.}}:
\batitle{A coherent photonic crossbar for scalable universal linear optics}.
\bjtitle{J. Light. Technol.}
\bvolume{41}(\bissue{8}),
\bfpage{2425}--\blpage{2442}
(\byear{2023})
\end{barticle}
\endbibitem

\bibitem[\protect\citeauthoryear{Rahimi~Kari et~al.}{2024}]{rahimi2024realization}
\begin{barticle}
\bauthor{\bsnm{Rahimi~Kari}, \binits{S.}},
\bauthor{\bsnm{Nobile}, \binits{N.A.}},
\bauthor{\bsnm{Pantin}, \binits{D.}},
\bauthor{\bsnm{Shah}, \binits{V.}},
\bauthor{\bsnm{Youngblood}, \binits{N.}}:
\batitle{Realization of an integrated coherent photonic platform for scalable matrix operations}.
\bjtitle{Optica}
\bvolume{11}(\bissue{4}),
\bfpage{542}--\blpage{551}
(\byear{2024})
\end{barticle}
\endbibitem

\bibitem[\protect\citeauthoryear{Moralis-Pegios et~al.}{2024}]{moralis2024perfect}
\begin{barticle}
\bauthor{\bsnm{Moralis-Pegios}, \binits{M.}},
\bauthor{\bsnm{Giamougiannis}, \binits{G.}},
\bauthor{\bsnm{Tsakyridis}, \binits{A.}},
\bauthor{\bsnm{Lazovsky}, \binits{D.}},
\bauthor{\bsnm{Pleros}, \binits{N.}}:
\batitle{Perfect linear optics using silicon photonics}.
\bjtitle{Nat. Commun.}
\bvolume{15}(\bissue{1}),
\bfpage{5468}
(\byear{2024})
\end{barticle}
\endbibitem

\bibitem[\protect\citeauthoryear{Tait et~al.}{2016}]{tait2016microring}
\begin{barticle}
\bauthor{\bsnm{Tait}, \binits{A.N.}},
\bauthor{\bsnm{Wu}, \binits{A.X.}},
\bauthor{\bsnm{De~Lima}, \binits{T.F.}},
\bauthor{\bsnm{Zhou}, \binits{E.}},
\bauthor{\bsnm{Shastri}, \binits{B.J.}},
\bauthor{\bsnm{Nahmias}, \binits{M.A.}},
\bauthor{\bsnm{Prucnal}, \binits{P.R.}}:
\batitle{Microring weight banks}.
\bjtitle{IEEE J. Sel. Top. Quantum Electron.}
\bvolume{22}(\bissue{6}),
\bfpage{312}--\blpage{325}
(\byear{2016})
\end{barticle}
\endbibitem

\bibitem[\protect\citeauthoryear{Tait et~al.}{2017}]{tait2017neuromorphic}
\begin{barticle}
\bauthor{\bsnm{Tait}, \binits{A.N.}},
\bauthor{\bsnm{De~Lima}, \binits{T.F.}},
\bauthor{\bsnm{Zhou}, \binits{E.}},
\bauthor{\bsnm{Wu}, \binits{A.X.}},
\bauthor{\bsnm{Nahmias}, \binits{M.A.}},
\bauthor{\bsnm{Shastri}, \binits{B.J.}},
\bauthor{\bsnm{Prucnal}, \binits{P.R.}}:
\batitle{Neuromorphic photonic networks using silicon photonic weight banks}.
\bjtitle{Sci. Rep.}
\bvolume{7}(\bissue{1}),
\bfpage{7430}
(\byear{2017})
\end{barticle}
\endbibitem

\bibitem[\protect\citeauthoryear{Feldmann et~al.}{2021}]{feldmann2021parallel}
\begin{barticle}
\bauthor{\bsnm{Feldmann}, \binits{J.}},
\bauthor{\bsnm{Youngblood}, \binits{N.}},
\bauthor{\bsnm{Karpov}, \binits{M.}},
\bauthor{\bsnm{Gehring}, \binits{H.}},
\bauthor{\bsnm{Li}, \binits{X.}},
\bauthor{\bsnm{Stappers}, \binits{M.}},
\bauthor{\bsnm{Le~Gallo}, \binits{M.}},
\bauthor{\bsnm{Fu}, \binits{X.}},
\bauthor{\bsnm{Lukashchuk}, \binits{A.}},
\bauthor{\bsnm{Raja}, \binits{A.S.}}, \betal:
\batitle{Parallel convolutional processing using an integrated photonic tensor core}.
\bjtitle{Nature}
\bvolume{589}(\bissue{7840}),
\bfpage{52}--\blpage{58}
(\byear{2021})
\end{barticle}
\endbibitem

\bibitem[\protect\citeauthoryear{Ohno et~al.}{2022}]{ohno2022si}
\begin{barticle}
\bauthor{\bsnm{Ohno}, \binits{S.}},
\bauthor{\bsnm{Tang}, \binits{R.}},
\bauthor{\bsnm{Toprasertpong}, \binits{K.}},
\bauthor{\bsnm{Takagi}, \binits{S.}},
\bauthor{\bsnm{Takenaka}, \binits{M.}}:
\batitle{Si microring resonator crossbar array for on-chip inference and training of the optical neural network}.
\bjtitle{ACS Photonics}
\bvolume{9}(\bissue{8}),
\bfpage{2614}--\blpage{2622}
(\byear{2022})
\end{barticle}
\endbibitem

\bibitem[\protect\citeauthoryear{Zhang et~al.}{2022}]{zhang2022silicon}
\begin{barticle}
\bauthor{\bsnm{Zhang}, \binits{W.}},
\bauthor{\bsnm{Huang}, \binits{C.}},
\bauthor{\bsnm{Peng}, \binits{H.-T.}},
\bauthor{\bsnm{Bilodeau}, \binits{S.}},
\bauthor{\bsnm{Jha}, \binits{A.}},
\bauthor{\bsnm{Blow}, \binits{E.}},
\bauthor{\bsnm{De~Lima}, \binits{T.F.}},
\bauthor{\bsnm{Shastri}, \binits{B.J.}},
\bauthor{\bsnm{Prucnal}, \binits{P.}}:
\batitle{Silicon microring synapses enable photonic deep learning beyond 9-bit precision}.
\bjtitle{Optica}
\bvolume{9}(\bissue{5}),
\bfpage{579}--\blpage{584}
(\byear{2022})
\end{barticle}
\endbibitem

\bibitem[\protect\citeauthoryear{Dong et~al.}{2023}]{dong2023higher}
\begin{barticle}
\bauthor{\bsnm{Dong}, \binits{B.}},
\bauthor{\bsnm{Aggarwal}, \binits{S.}},
\bauthor{\bsnm{Zhou}, \binits{W.}},
\bauthor{\bsnm{Ali}, \binits{U.E.}},
\bauthor{\bsnm{Farmakidis}, \binits{N.}},
\bauthor{\bsnm{Lee}, \binits{J.S.}},
\bauthor{\bsnm{He}, \binits{Y.}},
\bauthor{\bsnm{Li}, \binits{X.}},
\bauthor{\bsnm{Kwong}, \binits{D.-L.}},
\bauthor{\bsnm{Wright}, \binits{C.}}, \betal:
\batitle{Higher-dimensional processing using a photonic tensor core with continuous-time data}.
\bjtitle{Nat. Photon.}
\bvolume{17}(\bissue{12}),
\bfpage{1080}--\blpage{1088}
(\byear{2023})
\end{barticle}
\endbibitem

\bibitem[\protect\citeauthoryear{Ling et~al.}{2023}]{ling2023chip}
\begin{barticle}
\bauthor{\bsnm{Ling}, \binits{Q.}},
\bauthor{\bsnm{Dong}, \binits{P.}},
\bauthor{\bsnm{Chu}, \binits{Y.}},
\bauthor{\bsnm{Dong}, \binits{X.}},
\bauthor{\bsnm{Chen}, \binits{J.}},
\bauthor{\bsnm{Dai}, \binits{D.}},
\bauthor{\bsnm{Shi}, \binits{Y.}}:
\batitle{On-chip optical matrix-vector multiplier based on mode division multiplexing}.
\bjtitle{Chip}
\bvolume{2}(\bissue{4}),
\bfpage{100061}
(\byear{2023})
\end{barticle}
\endbibitem

\bibitem[\protect\citeauthoryear{Tang et~al.}{2024}]{tang2024symmetric}
\begin{barticle}
\bauthor{\bsnm{Tang}, \binits{R.}},
\bauthor{\bsnm{Ohno}, \binits{S.}},
\bauthor{\bsnm{Tanizawa}, \binits{K.}},
\bauthor{\bsnm{Ikeda}, \binits{K.}},
\bauthor{\bsnm{Okano}, \binits{M.}},
\bauthor{\bsnm{Toprasertpong}, \binits{K.}},
\bauthor{\bsnm{Takagi}, \binits{S.}},
\bauthor{\bsnm{Takenaka}, \binits{M.}}:
\batitle{Symmetric silicon microring resonator optical crossbar array for accelerated inference and training in deep learning}.
\bjtitle{Photonics Res.}
\bvolume{12}(\bissue{8}),
\bfpage{1681}--\blpage{1688}
(\byear{2024})
\end{barticle}
\endbibitem

\bibitem[\protect\citeauthoryear{Tang et~al.}{2022}]{tang2022two}
\begin{barticle}
\bauthor{\bsnm{Tang}, \binits{R.}},
\bauthor{\bsnm{Okano}, \binits{M.}},
\bauthor{\bsnm{Toprasertpong}, \binits{K.}},
\bauthor{\bsnm{Takagi}, \binits{S.}},
\bauthor{\bsnm{Englund}, \binits{D.}},
\bauthor{\bsnm{Takenaka}, \binits{M.}}:
\batitle{Two-layer integrated photonic architectures with multiport photodetectors for high-fidelity and energy-efficient matrix multiplications}.
\bjtitle{Opt. Express}
\bvolume{30}(\bissue{19}),
\bfpage{33940}--\blpage{33954}
(\byear{2022})
\end{barticle}
\endbibitem

\bibitem[\protect\citeauthoryear{Tang et~al.}{2024}]{tang2024single}
\begin{bchapter}
\bauthor{\bsnm{Tang}, \binits{R.}},
\bauthor{\bsnm{Okano}, \binits{M.}},
\bauthor{\bsnm{Toprasertpong}, \binits{K.}},
\bauthor{\bsnm{Takagi}, \binits{S.}},
\bauthor{\bsnm{Takenaka}, \binits{M.}}:
\bctitle{A single-wavelength non-coherent photonic matrix multiplication circuit for optical neural networks}.
In: \bbtitle{Conference on Lasers and Electro-Optics (CLEO)}
(\byear{2024})
\end{bchapter}
\endbibitem

\bibitem[\protect\citeauthoryear{Wuttig et~al.}{2017}]{wuttig2017phase}
\begin{barticle}
\bauthor{\bsnm{Wuttig}, \binits{M.}},
\bauthor{\bsnm{Bhaskaran}, \binits{H.}},
\bauthor{\bsnm{Taubner}, \binits{T.}}:
\batitle{Phase-change materials for non-volatile photonic applications}.
\bjtitle{Nat. Photon.}
\bvolume{11}(\bissue{8}),
\bfpage{465}--\blpage{476}
(\byear{2017})
\end{barticle}
\endbibitem

\bibitem[\protect\citeauthoryear{Zhou et~al.}{2023}]{zhou2023memory}
\begin{barticle}
\bauthor{\bsnm{Zhou}, \binits{W.}},
\bauthor{\bsnm{Dong}, \binits{B.}},
\bauthor{\bsnm{Farmakidis}, \binits{N.}},
\bauthor{\bsnm{Li}, \binits{X.}},
\bauthor{\bsnm{Youngblood}, \binits{N.}},
\bauthor{\bsnm{Huang}, \binits{K.}},
\bauthor{\bsnm{He}, \binits{Y.}},
\bauthor{\bsnm{David~Wright}, \binits{C.}},
\bauthor{\bsnm{Pernice}, \binits{W.H.}},
\bauthor{\bsnm{Bhaskaran}, \binits{H.}}:
\batitle{In-memory photonic dot-product engine with electrically programmable weight banks}.
\bjtitle{Nat. Commun.}
\bvolume{14}(\bissue{1}),
\bfpage{2887}
(\byear{2023})
\end{barticle}
\endbibitem

\bibitem[\protect\citeauthoryear{Miyatake et~al.}{2024}]{miyatake2024photonic}
\begin{barticle}
\bauthor{\bsnm{Miyatake}, \binits{Y.}},
\bauthor{\bsnm{Tang}, \binits{R.}},
\bauthor{\bsnm{Makino}, \binits{K.}},
\bauthor{\bsnm{Tominaga}, \binits{J.}},
\bauthor{\bsnm{Miyata}, \binits{N.}},
\bauthor{\bsnm{Okano}, \binits{M.}},
\bauthor{\bsnm{Toprasertpong}, \binits{K.}},
\bauthor{\bsnm{Takagi}, \binits{S.}},
\bauthor{\bsnm{Takenaka}, \binits{M.}}:
\batitle{Photonic matrix-vector multiplication with low-insertion-loss and non-volatile {Ge$_2$Sb$_2$Te$_3$S$_2$} intensity modulators}.
\bjtitle{J. Light. Technol.}
\bvolume{42}(\bissue{12}),
\bfpage{4347}--\blpage{4354}
(\byear{2024})
\end{barticle}
\endbibitem

\bibitem[\protect\citeauthoryear{Siew et~al.}{2021}]{siew2021review}
\begin{barticle}
\bauthor{\bsnm{Siew}, \binits{S.Y.}},
\bauthor{\bsnm{Li}, \binits{B.}},
\bauthor{\bsnm{Gao}, \binits{F.}},
\bauthor{\bsnm{Zheng}, \binits{H.Y.}},
\bauthor{\bsnm{Zhang}, \binits{W.}},
\bauthor{\bsnm{Guo}, \binits{P.}},
\bauthor{\bsnm{Xie}, \binits{S.W.}},
\bauthor{\bsnm{Song}, \binits{A.}},
\bauthor{\bsnm{Dong}, \binits{B.}},
\bauthor{\bsnm{Luo}, \binits{L.W.}}, \betal:
\batitle{Review of silicon photonics technology and platform development}.
\bjtitle{J. Light. Technol.}
\bvolume{39}(\bissue{13}),
\bfpage{4374}--\blpage{4389}
(\byear{2021})
\end{barticle}
\endbibitem

\bibitem[\protect\citeauthoryear{Ma et~al.}{2013}]{ma2013ultralow}
\begin{barticle}
\bauthor{\bsnm{Ma}, \binits{Y.}},
\bauthor{\bsnm{Zhang}, \binits{Y.}},
\bauthor{\bsnm{Yang}, \binits{S.}},
\bauthor{\bsnm{Novack}, \binits{A.}},
\bauthor{\bsnm{Ding}, \binits{R.}},
\bauthor{\bsnm{Lim}, \binits{A.E.-J.}},
\bauthor{\bsnm{Lo}, \binits{G.-Q.}},
\bauthor{\bsnm{Baehr-Jones}, \binits{T.}},
\bauthor{\bsnm{Hochberg}, \binits{M.}}:
\batitle{Ultralow loss single layer submicron silicon waveguide crossing for {SOI} optical interconnect}.
\bjtitle{Opt. Express}
\bvolume{21}(\bissue{24}),
\bfpage{29374}--\blpage{29382}
(\byear{2013})
\end{barticle}
\endbibitem

\bibitem[\protect\citeauthoryear{Wilkes et~al.}{2016}]{wilkes201660}
\begin{barticle}
\bauthor{\bsnm{Wilkes}, \binits{C.M.}},
\bauthor{\bsnm{Qiang}, \binits{X.}},
\bauthor{\bsnm{Wang}, \binits{J.}},
\bauthor{\bsnm{Santagati}, \binits{R.}},
\bauthor{\bsnm{Paesani}, \binits{S.}},
\bauthor{\bsnm{Zhou}, \binits{X.}},
\bauthor{\bsnm{Miller}, \binits{D.A.}},
\bauthor{\bsnm{Marshall}, \binits{G.D.}},
\bauthor{\bsnm{Thompson}, \binits{M.G.}},
\bauthor{\bsnm{O’Brien}, \binits{J.L.}}:
\batitle{60 {dB} high-extinction auto-configured {Mach--Zehnder} interferometer}.
\bjtitle{Opt. Lett.}
\bvolume{41}(\bissue{22}),
\bfpage{5318}--\blpage{5321}
(\byear{2016})
\end{barticle}
\endbibitem

\bibitem[\protect\citeauthoryear{}{}]{iris}
\begin{botherref}
https://www.kaggle.com/datasets/uciml/iris
\end{botherref}
\endbibitem

\bibitem[\protect\citeauthoryear{Gao et~al.}{2022}]{gao2022thermo}
\begin{barticle}
\bauthor{\bsnm{Gao}, \binits{F.}},
\bauthor{\bsnm{Xie}, \binits{W.}},
\bauthor{\bsnm{Tan}, \binits{J.Y.S.}},
\bauthor{\bsnm{Leong}, \binits{C.P.}},
\bauthor{\bsnm{Li}, \binits{C.}},
\bauthor{\bsnm{Luo}, \binits{X.}},
\bauthor{\bsnm{Lo}, \binits{G.-Q.}}:
\batitle{Thermo-optic phase shifter with interleaved suspended design for power efficiency and speed adjustment}.
\bjtitle{Micromachines}
\bvolume{13}(\bissue{11}),
\bfpage{1925}
(\byear{2022})
\end{barticle}
\endbibitem

\bibitem[\protect\citeauthoryear{}{}]{fashionMNIST}
\begin{botherref}
https://github.com/zalandoresearch/fashion-mnist
\end{botherref}
\endbibitem

\bibitem[\protect\citeauthoryear{Xu et~al.}{2021}]{xu202111}
\begin{barticle}
\bauthor{\bsnm{Xu}, \binits{X.}},
\bauthor{\bsnm{Tan}, \binits{M.}},
\bauthor{\bsnm{Corcoran}, \binits{B.}},
\bauthor{\bsnm{Wu}, \binits{J.}},
\bauthor{\bsnm{Boes}, \binits{A.}},
\bauthor{\bsnm{Nguyen}, \binits{T.G.}},
\bauthor{\bsnm{Chu}, \binits{S.T.}},
\bauthor{\bsnm{Little}, \binits{B.E.}},
\bauthor{\bsnm{Hicks}, \binits{D.G.}},
\bauthor{\bsnm{Morandotti}, \binits{R.}},
\bauthor{\bsnm{Mitchell}, \binits{A.}},
\bauthor{\bsnm{Moss}, \binits{D.J.}}:
\batitle{11 {TOPS} photonic convolutional accelerator for optical neural networks}.
\bjtitle{Nature}
\bvolume{589}(\bissue{7840}),
\bfpage{44}--\blpage{51}
(\byear{2021})
\end{barticle}
\endbibitem

\bibitem[\protect\citeauthoryear{Sludds et~al.}{2022}]{sludds2022delocalized}
\begin{barticle}
\bauthor{\bsnm{Sludds}, \binits{A.}},
\bauthor{\bsnm{Bandyopadhyay}, \binits{S.}},
\bauthor{\bsnm{Chen}, \binits{Z.}},
\bauthor{\bsnm{Zhong}, \binits{Z.}},
\bauthor{\bsnm{Cochrane}, \binits{J.}},
\bauthor{\bsnm{Bernstein}, \binits{L.}},
\bauthor{\bsnm{Bunandar}, \binits{D.}},
\bauthor{\bsnm{Dixon}, \binits{P.B.}},
\bauthor{\bsnm{Hamilton}, \binits{S.A.}},
\bauthor{\bsnm{Streshinsky}, \binits{M.}}, \betal:
\batitle{Delocalized photonic deep learning on the internet’s edge}.
\bjtitle{Science}
\bvolume{378}(\bissue{6617}),
\bfpage{270}--\blpage{276}
(\byear{2022})
\end{barticle}
\endbibitem

\bibitem[\protect\citeauthoryear{Chen et~al.}{2024}]{chen2024hypermultiplexed}
\begin{botherref}
\oauthor{\bsnm{Chen}, \binits{Z.}},
\oauthor{\bsnm{Ou}, \binits{S.}},
\oauthor{\bsnm{Xue}, \binits{K.}},
\oauthor{\bsnm{Zhou}, \binits{L.}},
\oauthor{\bsnm{Sludds}, \binits{A.}},
\oauthor{\bsnm{Hamerly}, \binits{R.}},
\oauthor{\bsnm{Zhang}, \binits{K.}},
\oauthor{\bsnm{Feng}, \binits{H.}},
\oauthor{\bsnm{Kopparapu}, \binits{R.}},
\oauthor{\bsnm{Lee}, \binits{C.}},
\oauthor{\bsnm{Zhong}, \binits{E.}},
\oauthor{\bsnm{Wang}, \binits{C.}},
\oauthor{\bsnm{Englund}, \binits{D.}},
\oauthor{\bsnm{Yu}, \binits{M.}}:
Hypermultiplexed integrated-photonics-based tensor optical processor.
Preprint at \url{https://doi.org/10.21203/rs.3.rs-4778342/v1}
(2024)
\end{botherref}
\endbibitem

\bibitem[\protect\citeauthoryear{Mojaver et~al.}{2024}]{mojaver2024recent}
\begin{botherref}
\oauthor{\bsnm{Mojaver}, \binits{K.R.}},
\oauthor{\bsnm{Safaee}, \binits{S.M.R.}},
\oauthor{\bsnm{Morrison}, \binits{S.S.}},
\oauthor{\bsnm{Liboiron-Ladouceur}, \binits{O.}}:
Recent advancements in mode division multiplexing for communication and computation in silicon photonics.
J. Light. Technol.,
1--12
(2024)
\doiurl{10.1109/JLT.2024.3412391}
\end{botherref}
\endbibitem

\bibitem[\protect\citeauthoryear{Huang et~al.}{2022}]{huang2022programmable}
\begin{barticle}
\bauthor{\bsnm{Huang}, \binits{Y.}},
\bauthor{\bsnm{Wang}, \binits{W.}},
\bauthor{\bsnm{Qiao}, \binits{L.}},
\bauthor{\bsnm{Hu}, \binits{X.}},
\bauthor{\bsnm{Chu}, \binits{T.}}:
\batitle{Programmable low-threshold optical nonlinear activation functions for photonic neural networks}.
\bjtitle{Opt. Lett.}
\bvolume{47}(\bissue{7}),
\bfpage{1810}--\blpage{1813}
(\byear{2022})
\end{barticle}
\endbibitem

\end{thebibliography}

\end{document}